\documentclass[%
 reprint,
 notitlepage,
superscriptaddress,
nofootinbib,
 amsmath,amssymb,
 aps,
twocolumn, 
]{revtex4-1}

\usepackage{graphicx}
\usepackage{dcolumn}
\usepackage{bm}

\usepackage{xcolor}
\usepackage{tabularx}
\usepackage[colorlinks]{hyperref}
\hypersetup{
    colorlinks,
    linkcolor={red!50!black},
    citecolor={blue!50!black},
    urlcolor={blue!80!black}
}
\usepackage{physics}
\usepackage{comment}
\usepackage{multirow}

\newcommand{\sixflav}[6]{(\{{#1},{#2}\}:\{#3,#4\}:\{#5,#6\})}
\newcommand{\sixflavor}[6]{\{{#1},{#2}\}:\{#3,#4\}:\{#5,#6\}}
\newcolumntype{Y}{>{\centering\arraybackslash}X}
\newcolumntype{Z}[1]{>{\centering\arraybackslash}m{#1}}

\begin{document}


\title{Probing neutrino production in high-energy astrophysical neutrino sources \\
with  the Glashow Resonance}

\author{Qinrui Liu}
\email{qinrui.liu@queensu.ca}
\affiliation{
Department of Physics, Engineering Physics and Astronomy,\\
Queen’s University, Kingston ON K7L 3N6, Canada
}
\affiliation{%
Arthur B. McDonald Canadian Astroparticle Physics Research Institute, Kingston ON K7L 3N6, Canada
}
\affiliation{Perimeter Institute for Theoretical Physics, Waterloo ON N2L 2Y5, Canada}

\author{Ningqiang Song}
\email{songnq@itp.ac.cn}
\affiliation{Institute of Theoretical Physics, Chinese Academy of Sciences, Beijing, 100190, China}
\affiliation{Department of Mathematical Sciences, University of Liverpool, Liverpool, L69 7ZL, United Kingdom}

\author{Aaron C. Vincent}
\email{aaron.vincent@queensu.ca}
\affiliation{
Department of Physics, Engineering Physics and Astronomy,\\
Queen’s University, Kingston ON K7L 3N6, Canada
}
\affiliation{%
Arthur B. McDonald Canadian Astroparticle Physics Research Institute, Kingston ON K7L 3N6, Canada
}
\affiliation{Perimeter Institute for Theoretical Physics, Waterloo ON N2L 2Y5, Canada}

\date{\today}

\begin{abstract}
The flavor composition of high-energy neutrinos carries important information about their birth. However, the two most common production scenarios, $pp$ (hadronuclear) and $p\gamma$ (photohadronic) processes, lead to the same flavor ratios when neutrinos and antineutrinos cannot be distinguished. The Glashow resonant interaction $\bar{\nu}_e+e^- \rightarrow W^-$ becomes a window to differentiate the antineutrino contribution from the total diffuse neutrino flux, thus lifting this degeneracy.  We examine the power of Glashow resonant events in measuring the fraction of the $\bar{\nu}_e$ flux with current IceCube data, and produce projected sensitivities based on the combined exposure of planned Cherenkov neutrino telescopes around the globe. We find that $pp$ and $p\gamma$ can be distinguished at a 2$\sigma$ significance level in the next decades, in both an event-wise analysis and a more conservative statistical analysis, even with pessimistic assumptions on the spectral index of the astrophysical flux. Finally, we consider the sensitivity of future experiments to mixed production mechanisms. 

\end{abstract}

\maketitle


\section{Introduction}

\renewcommand{\arraystretch}{1.4}
\begin{table*}[hbtp!]
    \centering
    \begin{tabularx}{\textwidth}{Z{2.5cm}Z{3cm}Z{4cm}Z{6cm}Z{2.5cm}}
    \hline
    \centering
    Production  & Source flavor ratio & Earth flavor ratio $\nu+\bar{\nu}$ & Earth flavor ratio & $f_{\bar{\nu}_e}$  \\
    \hline
    $pp$  & $\sixflavor{1}{1}{2}{2}{0}{0}$ & $0.33:0.34:0.33$ & $\sixflavor{0.17}{0.17}{0.17}{0.17}{0.16}{0.16}$ & 0.17 \\
    $pp\,\mu$ damped & $\sixflavor{0}{0}{1}{1}{0}{0}$  & $0.23:0.39:0.38$ &  $\sixflavor{0.11}{0.11}{0.20}{0.20}{0.19}{0.19}$ & 0.11 \\
    $p\gamma$ & $\sixflavor{1}{0}{1}{1}{0}{0}$  & $0.33:0.34:0.33$ &  $\sixflavor{0.26}{0.08}{0.21}{0.13}{0.20}{0.13}$ & 0.08 \\
    $p\gamma\,\mu$ damped &  $\sixflavor{0}{0}{1}{0}{0}{0}$ & $0.23:0.39:0.38$ & $\sixflavor{0.23}{0.00}{0.39}{0.00}{0.38}{0.00}$ & 0\\
    \hline    
    \end{tabularx}
    \caption{The composition of high-energy astrophysical neutrinos at the source and Earth arising from different production mechanisms. The flavor ratios are shown as \sixflavor{$f_{\nu_e}$}{$f_{ \bar{\nu}_e}$}{$f_{\nu_\mu}$}{$f_{\bar{\nu}_\mu}$}{$f_{{\nu}_\tau}$}{$f_{\bar{\nu}_\tau}$}, except for $\nu+\bar{\nu}$, where the contribution from neutrinos and antineutrinos are summed over. The standard neutrino oscillation effect is applied. The fraction of $\bar{\nu}_e$ flux in the total neutrino flux at Earth is displayed in the last column.
    }
    \label{tab:flavor_ratio}
\end{table*}

It has been a decade since the discovery of high-energy astrophysical neutrinos of TeV-PeV at the IceCube Observatory~\cite{IceCube:2013low}. So far, only two individual sources have revealed themselves~\cite{IceCube:2018dnn,IceCube:2018cha,IceCube:2022der}, and the origin and production of the diffuse flux of neutrinos remains opaque. 

One promising handle on the origin of this flux lies in the flavor composition of the incident neutrinos~\cite{Rachen:1998fd,Athar:2000yw,Crocker:2001zs,Barenboim:2003jm,Beacom:2003nh,Beacom:2004jb,Kashti:2005qa,Mena:2006eq,Kachelriess:2006fi,Lipari:2007su,Esmaili:2009dz,Choubey:2009jq,Hummer:2010ai,Palladino:2015zua,Bustamante:2015waa,Biehl:2016psj,Bustamante:2019sdb,Beacom:2002vi,Barenboim:2003jm,Beacom:2003nh,Beacom:2003eu,Beacom:2003zg,Serpico:2005bs,Mena:2006eq,Lipari:2007su,Pakvasa:2007dc,Esmaili:2009dz,Choubey:2009jq,Esmaili:2009fk,Bhattacharya:2009tx,Bhattacharya:2010xj,Bustamante:2010nq,Mehta:2011qb,Baerwald:2012kc,Fu:2012zr,Pakvasa:2012db,Chatterjee:2013tza,Xu:2014via,Aeikens:2014yga,Arguelles:2015dca,Bustamante:2015waa,Pagliaroli:2015rca,deSalas:2016svi,Gonzalez-Garcia:2016gpq,Bustamante:2016ciw,Rasmussen:2017ert,Dey:2017ede,Bustamante:2018mzu,Farzan:2018pnk,Ahlers:2018yom,Brdar:2018tce,Palladino:2019pid,Ahlers:2020miq,Karmakar:2020yzn,Fiorillo:2020gsb}. Even though neutrino oscillation brings this ratio near equality, the structure of the PMNS mixing matrix means that there remains an imprint of the flavor composition at the source. Reconstructing the flavors of these neutrinos can yield valuable information on the production mechanism, propagation, and physics in the detector. Previous work has examined the flavor composition of the IceCube neutrinos in great detail~\cite{Mena:2014sja,Chen:2014gxa,Palomares-Ruiz:2015mka,Palladino:2015zua,Bustamante:2015waa,Vincent:2016nut,Palladino:2019pid,Song:2020nfh,Aartsen:2015ivb,Aartsen:2015knd,IceCube:2020fpi,IceCube:2021tdn}, including in the context of new physics (e.g.~\cite{Arguelles:2015dca,Ahlers:2018yom,Arguelles:2019tum,Ahlers:2020miq,Gonzalez-Garcia:2016gpq,Mack:2019bps,Song:2020nfh}). 

The majority of these studies have focused on the 3-flavor composition assuming there is an equal contribution from neutrinos and antineutrinos. However, it is critical to note that different neutrino production scenarios, as well as potential new physics effects in propagation, encode information in a six-flavor parameter space that includes neutrinos and antineutrinos. As high-energy sources mainly accelerate protons, an asymmetry between particles and antiparticles is expected, and the two most credible astrophysical neutrino production mechanisms, hadronuclear ($pp$) and photohadronic ($p \gamma$) interactions, yield different $\pi^+\pi^-$ ratios after hadronization, and therefore different $\bar \nu$ to $\nu$ ratios.

A Cherenkov neutrino telescope like IceCube uses natural ice or water as a detector medium. Optical modules embedded in the medium are triggered by the Cherenkov photons emitted by relativistic charged particles produced by charged-current (CC) or neutral-current (NC) deep inelastic scattering (DIS) of high-energy neutrinos with nucleons. Flavors can be inferred on a statistical basis, using the charged secondaries produced in these interactions. However, as the valence quark contribution to nucleons becomes negligible at high energies, it becomes practically impossible to separate neutrinos from antineutrinos~\cite{Gandhi:1995tf,Formaggio:2013kya}. 

 Fortunately, owing to the abundance of electrons on Earth, the resonant scattering around the $W$ pole~\cite{Glashow:1960zz}, also called Glashow resonance (GR), i.e. ${\bar{\nu}_{e} + e^{-} \rightarrow W^{-}_{} \rightarrow \text{anything}}$, dramatically enhances the scattering cross section near the neutrino energy of 6.3~PeV. This provides an avenue to tease out the $\bar{\nu}_e$ flux from the total neutrino flux. The detectability of GR events at a high-energy neutrino telescope such as IceCube and the possibility of diagnosing high-energy astrophysical processes for the benchmark cases and more complex scenarios have been extensively studied~\cite{Anchordoqui:2004eb,Hummer:2010ai,Bhattacharya:2011qu,Xing:2011zm,Bhattacharya:2012fh,Barger:2012mz,Anchordoqui:2014yva,Barger:2014iua,Palladino:2015uoa,Anchordoqui:2016ewn,Biehl:2016psj,Huang:2019hgs,Fiorillo:2022rft,Huang:2023yqz}. The potential of constraining new physics such as neutrino decay and Lorentz invariance violation has also been discussed~\cite{Shoemaker:2015qul,Bustamante:2020niz,Xu:2022svm}. Such event had not been  observed until recently where IceCube published the observation of the first GR candidate event, a partially contained cascade event with a visible energy of 6.05$\pm$0.72~PeV, at a 2.3$\,\sigma$ significance assuming a $E^{-2.5}$ spectrum~\cite{IceCube:2021rpz}. This is an event passing the selection of PeV energy partially-contained events (PEPE)~\cite{Lu:2017nti}.  This observation and the non-observation of GR events with other IceCube event selections make it possible for us to study its implications for processes in neutrino sources. A very recent study of GR detection from the $\nu_e + \bar{\nu}_e$ flux~\cite{Huang:2023yqz} performed an analysis with current IceCube observation focusing on shower events in the GR energy window, and excluded the muon damped $p\gamma$ scenario at $2\sigma$, whilst showing that  IceCube-Gen2 may be able to detect $pp$ type sources in next decades.
 
With a number of next-generation neutrino telescopes in the future, we are able to study the prospects of the detection of such events and what they can infer with the optimized power in the upcoming years. Here, with a special focus on the asymmetries between the parent $\pi^+$ and $\pi^-$, we will ask how much information can be gained from the GR in light of upcoming data from neutrino telescopes under construction in the next decades. Specifically, we will combine the questions of collision mechanisms ($pp$ versus $p\gamma$), with neutrino production: direct pion decay, or muon-damped pion decay in the presence of a strong magnetic field. 
 We consider the IceCube public data, as well as the potential of future neutrino telescopes.
To find the sensitivity of future telescopes, we combine the forecasted sensitivities of Baikal-GVD~\cite{Baikal-GVD:2018isr}, IceCube-Gen2~\cite{IceCube-Gen2:2020qha}, KM3Net~\cite{KM3Net:2016zxf}, P-ONE~\cite{P-ONE:2020ljt} and TRIDENT~\cite{Ye:2022vbk}.

In this work, we explore two scenarios pertinent to the possible identification of GR events. In the optimistic case, GR events can be distinguished through the event cuts involving morphology, energy deposition, muons from hadronic decay, etc~\cite{IceCube:2021rpz}. In the modest case, GR events are indistinguishable from other DIS events, and the analysis is done based on statistics alone. 

It is also likely that both $pp$ and $p\gamma$ contribute significantly to the source of high-energy astrophysical neutrinos considering that there is more than one source population. The investigation of GR helps pin down the fractional contribution of each production mechanism. More broadly speaking, the more realistic hadronuclear and photohadronic interactions at high energies predict more complicated neutrino production pictures than the reference $pp$ and $p\gamma$ scenarios, which is reflected in the relative numbers of $\pi^+$ and $\pi^-$. The study of the $\pi^+/\pi^-$ ratio provides important insights into the physical processes in the source.

This paper is structured as follows. We will start with a general description of neutrino sources in Sec.~\ref{sec:source}. In Sec.~\ref{sec:framework} we will introduce our framework. Then we will investigate the identification of GR events on the event-by-event basis with its observation prospects and implications in Sec.~\ref{sec:GR_ID}. We also perform a study on the statistical basis where we examine the 4-flavor composition of the neutrinos, i.e. 3-flavor plus $\bar{\nu}_e$ in Sec.~\ref{sec:stats}, which gives a coherent and complete picture of information we can learn from flavor compositions. Furthermore, in Sec.~\ref{sec:combination} we explore the sensitivity of differentiating a mixed contribution from different production mechanisms. We conclude in Sec.~\ref{sec:conclusion}.

\section{High-energy Neutrino Sources}
\label{sec:source}
Cosmic rays produced in cosmic accelerators inevitably interact with surrounding matter or radiation, predominantly producing  pions which decay into neutrinos via $\pi^{+/-}\rightarrow \mu^{+/-} + \nu_\mu / \bar{\nu}_\mu$. The muons further decay via the process $\mu^{+/-}\rightarrow e^{+/-} + \nu_e/\bar{\nu}_e + \bar{\nu}_\mu/\nu_\mu$. Neutrino source candidates have been widely discussed, predominantly extragalactic suggested by the results of IceCube all-sky scans~\cite{IceCube:2019cia,IceCube:2022der}.
Photohardonic processes are typically expected in jets or/and cores of active galactic nuclei (AGN) ~\cite{Mannheim:1995mm,Halzen:1997hw,Atoyan:2001ey,Stecker:1991vm,Alvarez-Muniz:2004xlu} and jets of Gamma-ray burst~\cite{Waxman:1997ti,Dermer:2003zv,Murase:2006mm} while for starburst galaxies (SBGs)~\cite{Thompson:2006np,Murase:2013rfa,Tamborra:2014xia}, hadronuclear reactions are anticipated to be the main contribution.  
Nevertheless, the true case can be more complicated depending on the geometry of sources. For the two sources in directions showing neutrino flux excess beyond 3$\sigma$, TXS 05060+056~\cite{IceCube:2018dnn,IceCube:2018cha} - a jetted AGN and NGC 1068~\cite{IceCube:2022der} - a non-jetted AGN with a high level of star formation, which belong to different source catalogs and are expected to have distinct neutrino production mechanisms where the leading energy range differs, models on both $pp$ and $p\gamma$ production have been discussed with multi-wavelengths observations, e.g. in~\cite{Keivani:2018rnh,Murase:2018iyl,Gao:2018mnu,Liu:2018utd,Reimer:2018vvw,Murase:2019vdl,Inoue:2019yfs,Kheirandish:2021wkm,Eichmann:2022lxh}. It is crucial to discriminate $pp$ and $p\gamma$ scenarios to determine the neutrino production in neutrino sources. These two processes both lead to the production of pions but with different $\pi^+/\pi^-$ ratios. In the $p\gamma$ scenario, the dominant process resulting in neutrino production is $p+\gamma \rightarrow \Delta^+\rightarrow \pi^+ + n$. This neutron can escape the source, but the energy carried by the neutrino from its decay will be much lower than those produced by the pion. At the same time, $pp$ interactions give a more uniform distribution of pion charges $p+p\rightarrow n_\pi\left [\pi^0 +\pi^+ + \pi^- \right]$ where $n_\pi$ is the multiplicity factor.

In a 3-flavor analysis, both scenarios give the pion decay produced neutrino flavor ratio $(f_{e}:f_{\mu}:f_{\tau})_s = (1:2:0)$, where the subscript $s$ refers to the source. We write the 6-flavor composition as:
$\sixflav{f_{\nu_e}}{f_{ \bar{\nu}_e}}{f_{\nu_\mu}}{f_{\bar{\nu}_\mu}}{f_{{\nu}_\tau}}{f_{\bar{\nu}_\tau}}$. 
Accounting for the asymmetry in charges, this  gives $\sixflav{1}{0}{1}{1}{0}{0}$ for the $p\gamma$ scenario and $\sixflav{1}{1}{2}{2}{0}{0}$ for the $pp$ scenario, breaking the degeneracy. If a very strong magnetic field is present in the source, muons in the decay chain could lose energy significantly via synchrotron radiation, and the consequent decay would be suppressed. This is the \textit{muon damped} scenario, which removes the electron (anti)neutrino component, and thus  no antineutrinos are expected for the $p\gamma$, muon-damped scenario. In addition to the pion decay scenarios, there may be sources generating a pure $\bar{\nu}_e$ flux from a neutron beam \cite{Anchordoqui:2003vc}. Protons and neutrons are produced from photodisintegration of Fe by background photon fields. While protons can rapidly lose energy or get deflected by the magnetic field, the neutrons would decay in flight leading to a pure $\bar \nu_e$ signal. However, we will not delve into details of this scenario as it is expected to contribute to the neutrino flux with energies lower than the GR energy as a neutrino from neutron decay only takes away a small fraction of the neutron energy and this scenario is disfavored by current observation for the large $\bar{\nu}_e$ fraction \cite{IceCube:2020fpi}.

Table.~\ref{tab:flavor_ratio} shows a summary of the flavor ratios at the source for the scenarios discussed above. As neutrinos travel from the source to Earth, the flavor composition changes as a consequence of neutrino oscillation, which yields a more uniform composition; however, in the standard oscillation picture, neutrino-antineutrino mixing does not occur. In this work, we use the best-fit oscillation parameters from NuFIT~5.1~\cite{Esteban:2020cvm} with normal mass ordering. The fraction of $\bar{\nu}_e$ flux in the total neutrino flux at Earth, $f_{\bar{\nu}_e}$ is also listed. Going forward, we will focus on the most plausible scenarios involving pion decays. We will not consider the contribution of neutrino flux from heavier hadrons (e.g. kaons and charmed hadrons), whose decay products are expected to be very subdominant~\cite{Hummer:2010ai,Winter:2012xq,Biehl:2016psj}.

\section{Analysis Framework}
\label{sec:framework}

The cross section for GR reads 
\begin{align}
\sigma^{}_{\bar{\nu}^{}_{e}e}(s) 
=\,& 24\pi\,\Gamma^2_W\,{\rm Br}(W^-{\rightarrow}\,\overline{\nu}_e+e^-) \nonumber\\
&\hspace{0cm}\times \frac{s/M^2_W}{(s-M^2_W)^2+(M_W\Gamma_W)^2} \;,
\label{eq:xsec}
\end{align}
where $\sqrt{s}$ is the center-of-mass energy, $\Gamma^{}_{W}$ is the total decay rate of $W$, and ${\rm Br}(W^-\rightarrow\overline{\nu}_e+e^-) \simeq 10.7\% $ is the branching ratio of the decay channel $W^-\rightarrow\overline{\nu}_e+e^-$. The cross section $\sigma^{}_{\overline{\nu}^{}_{e}e}(s)$ reaches its maximum value $\sigma^{\rm GR}_{\overline{\nu}^{}_{e}e} \simeq 4.86 \times 10^{-31}~{\rm cm^2}$ when $\sqrt{s} = M^{}_{W} \simeq 80.4~{\rm GeV}$ \cite{Workman:2022ynf}. Since the target electrons are effectively at rest, the resonance peaks at a neutrino energy of $E^{}_{\overline{\nu}_{e}} \simeq 6.3~{\rm PeV}$. We do not include the effects on the cross section from the initial state radiation and Doppler broadening discussed in~\cite{Huang:2023yqz} as they are not included in the public IceCube Monte Carlo. This correction is expected to give rise to a lower peak at the resonance energy and higher tail at energies above, but we expect that it will not lead to a significant effect on our analyses due to the rapidly-falling power-law neutrino flux.

In a Cherenkov neutrino telescope, there are three event \textit{morphologies}, depending on the incoming neutrino flavor and interaction process. A \textit{cascade} is induced by an electron neutrino CCDIS or NCDIS of a neutrino of any flavor. A muon neutrino CCDIS would give a \textit{track} and a tau neutrino with an energy $\gtrsim$~PeV CCDIS can yield  a \textit{double-cascade} if the tau decays far enough from the primary interaction vertex, but still inside the detector. For GR events, the main $W$ decay channel is to hadrons, with a $\simeq 67\%$ probability, resulting in a cascade. Decays to each of the charged leptons $\ell^-$ occur $\simeq 11\%$ of the time; this will most likely yield a track-type event for decays to $\ell^- = \mu^-$, and a cascade if $\ell^-=e$ or $\tau$. Double-cascades are not expected to occur here, since a shower is not expected at the primary vertex.

The number of events expected at a neutrino detector is
\begin{align}
\begin{split}
  & N(E_{dep},\theta_z) \\ & =   T_{live}\int A_{eff}(E_\nu, \theta_z) \phi(E_\nu)\mathcal{P}\left (E_{dep}, E_\nu \right)dE_\nu d\Omega, 
\end{split}
\label{eq:event_rate}
\end{align}
where $\phi(E_\nu)\propto E^{-\gamma}_\nu$ is the diffuse astrophysical neutrino spectrum which we assume following a power-law spectrum and $T_{live}$ is the livetime for the data taking. $\mathcal{P}$ is the probability of depositing energy $E_{dep}$ for a neutrino of energy $E_\nu$, and takes into account the event reconstruction. $\theta_z$ is the zenith angle of the flux. $A_{eff}(E_\nu, \theta_z) = A_{geo}(\theta_z)\left[1 - \mathrm{exp}\left(-l(\theta_z)/\lambda_\nu(E_\nu)\right)\right]$ is the effective area for detecting neutrino events, where $A_{geo}$ is the geometric area of the instrumented volume
facing the incident neutrino and $l$ is the path length. $\lambda_\nu$ is the mean free path of neutrinos determined by the target density and cross section. For GR interactions, $\lambda^{GR}_{\bar{\nu}_e}= \left[n_e\sigma_{\bar{\nu}_e\,e} \right]^{-1}$ and for neutrino-nucleon DIS events, $\lambda^{CC/NC}_{\nu}= \left[n_{p/n}\sigma^{CC/NC}_{\nu,\, p/n} \right]^{-1}$ which applies to all neutrino flavors where $n_x$ the number density of electrons or nucleons.

Here, we conduct our study using the high-energy starting events (HESE)~\cite{IceCube:2020wum} and its future projection which include contained neutrino events above 60~TeV. We use the publicly-relaeased 7.5yr HESE data and Monte Carlo (MC)\footnote{\url{https://github.com/icecube/HESE-7-year-data-release/}}, which implicitly contains efficiency, effective area, angular and detector response information, including the effects of event reconstruction such as energy loss and misidentification that may affect our analysis.

To extend our analysis to projected future sensitivities, we still use the HESE MC, as it represents the most accurate parametrisation of the response of a large optical neutrino telescope. For these projections, we scale exposure (and thus expected number of events) proportionately to the volume $V$ and livetime $T_{live}$ of future detectors, i.e. the IceCube effective livetime $T_{eff} \simeq T_{live}(V/1\mathrm{km^3})$ where 1$\mathrm{km^3}$ corresponds to the IceCube volume., so that Eq.~\eqref{eq:event_rate} still holds by replacing $T_{live}$ with $T_{eff}$. 

In this work, we explore the potential of future neutrino telescopes that will be deployed in the next decade. \textbf{IceCube-Gen2}~\cite{IceCube-Gen2:2020qha} will extend the current IceCube detector to a volume of 8$\mathrm{km^3}$, by installing 120 additional strings. The high-energy module of \textbf{KM3NeT}~\cite{KM3Net:2016zxf} will be instrumented with two 115-string arrays, which in combination provides an effective volume of 2.8$\mathrm{km^3}$. Although \textbf{Baikal-GVD} has been taking data since 2018~\cite{Baikal-GVD:2018isr}, the detector is expected to be upgraded over time, and reach a final instrumented volume of 1.5$\mathrm{km^3}$ consisting of 90 strings. \textbf{P-ONE}~\cite{P-ONE:2020ljt}, a planned water-Cherenkov detector as KM3NeT and Baikal-GVD, will be constructed as a cylindrical configuration using 70 strings, with an estimated volume of 3.2$\mathrm{km^3}$. Here, we also include \textbf{TRIDENT}~\cite{Ye:2022vbk}, which will be deployed in South China Sea, 180~km away from the Yongxing Island. 1211 stings of 0.7~km length, each instrumented with 20 DOMs (digital optical modules), will distribute following the pattern of the Penrose tiling. This yields a total geometric volume of 7.5$\mathrm{km^3}$. Trident has finished the pathfinder program to test the oceanographic conditions and the detector performance~\cite{Ye:2022vbk}, and the first stage of the experiment is also funded.

We assume that Baikal-GVD and KM3Net will start taking data in 2025 and IceCube-Gen2, P-ONE and TRIDENT will be turned on in 2030. We will show our projection to 2040 by scaling up the effective area of IceCube, which corresponds to a 10-year exposure of all underground or underwater Cherenkov high-energy neutrino telescopes \cite{Song:2020nfh}.

For through-going tracks, as it is challenging to observe GR events in CC tracks mostly due to the subdominant decay branching ratio and the limitation of the energy reconstruction~\cite{Huang:2019hgs}, contained events or partially contained ones which can be reconstructed well, are still expected to contribute dominantly. As the resonant events happen at $\sim$PeV energies, the expected number of events is also very sensitive to the spectral index of the astrophysical flux. In characterizations of the diffuse astrophysical flux at IceCube, analyses of through-going muons give the hardest spectrum, $\gamma=2.37$ \cite{IceCube:2021uhz} while HESE yields the softest spectrum $\gamma=2.87$ \cite{IceCube:2020wum}. Analyses with cascade-only events or combined samples fall between the two \cite{Aartsen:2015knd,IceCube:2020acn}. In this work, we perform the analysis based on two spectral assumptions reflecting the muon and HESE numbers, which we respectively label as optimistic ($\gamma=2.37$) hard spectrum and conservative ($\gamma=2.87$) soft spectrum.

We will examine two separate scenarios: one in which GR events can be identified on an \textit{event-by-event} basis, and a scenario where the GR event rate can only be measured \textit{statistically}, which we will discuss in the next two sections. 

\section{Event-wise GR identification}
\label{sec:GR_ID}

In this section, we consider the case where we are able to identify GR events on an \textit{event-by-event} basis. 
We will start by describing how GR events can be singled out. We then determine what  current observations tell us about the electron antineutrino fraction, before moving on to  the future detection prospects and implications. 

A cascade is expected from the dominant hadronic decay channel of the $W$ ( $W\rightarrow$ hadrons), as well as from the leptonic channels $W^- \rightarrow \tau^{-}+\bar{\nu}_\tau$ and $W^- \rightarrow e^{-}+\bar{\nu}_e$.  The  hadronic  channel gives rise to a distinct signal: meson decays produce muons with energy of tens of GeV, which travel faster in-medium than Cherenkov photons and are visible as early pulses~\cite{IceCube:2021rpz}. This distinguishes them from the showers in CCDIS cascades. The only irreducible background here is cascades induced by NC all-flavor DIS. NCDIS showers are purely hadronic, but in such events the final-state neutrino carries away a large fraction of energy. This means that for the same deposited energy, the incoming neutrino energy is required to be much larger to mimic a GR event. Although this contribution is irreducible, the power-law flux naturally suppresses it.

In contrast, the cascade events induced by the two leptonic channels ($e$ and $\tau$) are unlikely to be differentiated from a cascade induced by DIS, as they will look functionally very similar to DIS CC events.

The $W^-\rightarrow \mu^{-} + \bar{\nu}_\mu$ channel would give a track without a cascade at the vertex, potentially differentiating it from CC $\nu_\mu$ DIS events which start with a hadronic cascade. For our future projections, we will assume that such an identification is plausible. In our event-wise analysis, we thus assume that the hadronic and $\mu^{-} +\bar{\nu}_\mu$ channels of $W^-$ decay can be perfectly differentiated from DIS events. Here, the hadronic channel still plays the major role due to the large branching ratio. 

The PEPE selection results in a factor of $\sim2$ increase in the effective area near the GR energy~\cite{IceCube:2021rpz}. To account for this, we will compare the HESE event rate with an effective PEPE event rate, which is a factor of two larger.

\begin{figure}[ht!]
    \hspace{-.8cm}
    \includegraphics[scale=0.5]{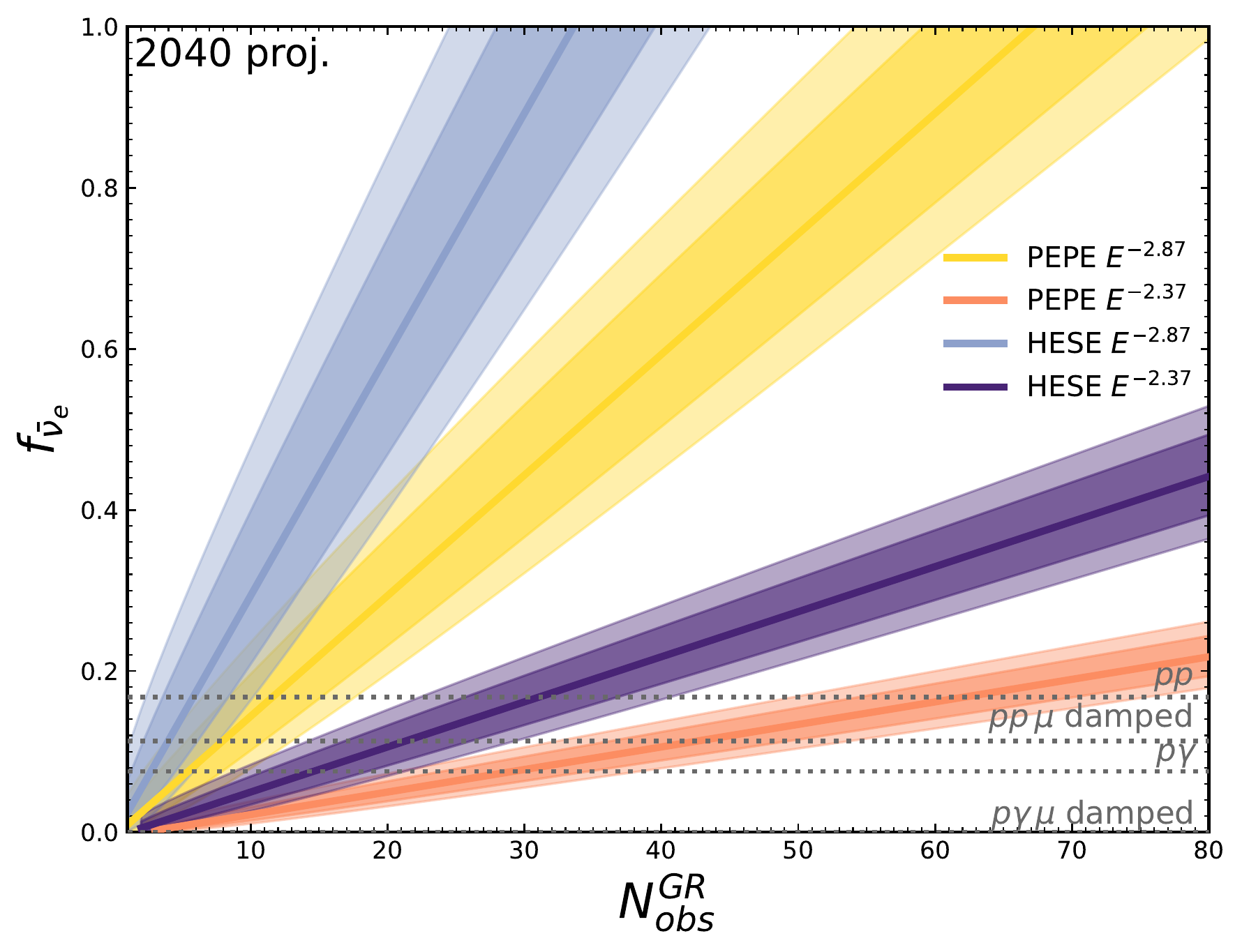}
    \caption{68\% (darker) and 90\% (lighter) C.L. intervals of the fraction $\bar{\nu}_e$ flux in the total high-energy astrophysical neutrino flux at Earth $f_{\bar{\nu}_e}$ as a function of the GR event observed by 2040, combining the detection in future neutrino telescopes. The solid lines show the best-fit $f_{\bar{\nu}_e}$ for each number of observed GR-like events. We show two spectral indices $\gamma=2.37$ and $\gamma=2.87$, with HESE selection and PEPE selection, respectively. See text for details. The gray dotted lines represent the $\bar{\nu}_e$ fractions predicted by the standard neutrino production mechanisms discussed, as labeled in the figure.}
    \label{fig:identified_GR}
\end{figure}

This GR candidate event observed by IceCube is a cascade in PEPE but is not in the HESE sample. Both of these facts provide constraints on the $\bar \nu_e$ flux. Ref.~\cite{IceCube:2021rpz} indicates that this event is consistent with a hadronic shower from $W$ decay. We set a deposited energy window $\in \left [4,10\right ]$~PeV, and with the event rates computed with the HESE MC, we find that observation of 1 event in 4.6 years of livetime with PEPE selection constrains 
\begin{equation}
  2\% \le  f_{\bar{\nu}_e} \le 72\%,
\end{equation}
at Earth assuming the hard spectrum and 
 \begin{equation}
     f_{\bar{\nu}_e} \gtrsim 10\%,
 \end{equation}
 assuming the soft spectrum at 90\% confidence level (C.L.), using Feldman-Cousins intervals~\cite{Feldman:1997qc}. 

Following the same procedure, non-observation of 1 event in the 7.5yr livetime with the HESE selection constrains 
\begin{equation}
   f_{\bar{\nu}_e} \le 51\%,
\end{equation}
with the hard spectrum. The soft spectrum is too soft to yield a constraint from HESE. These limits are consistent with our expectation that the high-energy neutrino flux dominated by muon damped $p\gamma$ source is disfavored with the observation of one GR event.

We now ask how well the $\bar{\nu}_e$ fraction can be constrained as a function of the number of observed GR events, and how this translates to the observation in future telescopes. We then examine the case where there is a good identification of GR events and contrast the PEPE selection with the HESE selection. We also keep the deposited energy window $\in \left [4,10\right ]$~PeV where GR events dominate and are likely to be distinguishable from other DIS events, and split it to 3 energy bins as in~\cite{IceCube:2021rpz}. The likelihood function for the $\bar \nu_e$ fraction can be written as

\begin{align}
\begin{split}
  & \mathcal{L}^{GR}(f_{\bar{\nu}_e})= \prod_i \mathcal{L}_{cas,i}\prod_j\mathcal{L}_{tr,j} 
  \\& =  e^{-N_{cas} -N_{tr} } \prod_i \frac{N^{N^{obs}_{cas,\,i}}_{cas,\,i}}{N^{obs}_{cas,\,i}!} \prod_j \frac{N_{tr,\,j}^{N^{obs}_{tr,\,j}}}{N^{obs}_{tr,\,j}!},   
   \label{eq:likelihood_GR}
\end{split}
\end{align}
where $N_{cas} = N^{GR}_{cas}+ N^{NC,\,\nu/\bar{\nu}}$ and $N_{tr} = N^{GR}_{tr}$ are expected GR-like cascade events and track events, with the superscript $obs$ used to label the observed number of events. As a result, one can obtain the $\bar{\nu}_e$ flux and then infer $f_{\bar{\nu}_e}$ by comparing this flux to the measured diffuse all-flavor flux. 

We apply our event-wise analysis to the future detection. We include the irreducible background from NCDIS events as the expected event number can reach above 1 event. The effect of including such background is negligible if we expect to observe a decent amount of GR events. However, if $f_{\bar{\nu}_e}$ is small, the NCDIS background will be important in the analysis. We then compute the intervals with the Asimov data\footnote{In the Asimov data set, the number of events in each bin is set to the expectation value in that bin.}, generated using Eq.~\eqref{eq:likelihood_GR} and the HESE MC. 

Fig.~\ref{fig:identified_GR} shows the expected constraints on $f_{\bar{\nu}_e}$ by 2040 depending on the number of total GR-like events we will observe in the energy window with all future neutrino telescopes, i.e. $N^{GR}_{obs}=\sum_i N^{obs}_{cas,\,i} + \sum_j N^{obs}_{tr,\,j}$. We set the all-flavor spectrum fixed to the soft or hard spectrum assumption and the only free parameter is $f_{\bar{\nu}_e}$. The uncertainties on $f_{\bar{\nu}_e}$ shrink with a harder neutrino spectrum and a larger $f_{\bar{\nu}_e}$, as more GR events are to be expected in neutrino telescopes. In the regime of pion decay and muon damped pion decay, $f_{\bar{\nu}_e}$ is always expected to be lower than 0.2. 

In this range, the widths of the 68\% CL bands vary from between $0.03$ to $0.1$. The special scenario that $f_{\bar{\nu}_e} = 0 $, has an uncertainty that is $\lesssim 1\%$, which makes the muon-damped $pp$ scenario easy to differentiate from muon-damped $p\gamma$ at the 5$\sigma$ level, even under our pessimistic spectrum and selection assumptions.  The significance to which each scenario could be distinguished is summarized in Table~\ref{tab:significance}. It is clear that  GR events will help in differentiating the degenerate 3-flavor  scenarios in the future. When the spectrum is hard, we will be confident to disentangling the scenarios while for the soft spectrum, we lose sensitivity but can still obtain $\simeq 2\sigma$.  

\begin{figure*}[htbp!]
    \centering
    \includegraphics[width=1.0\textwidth]{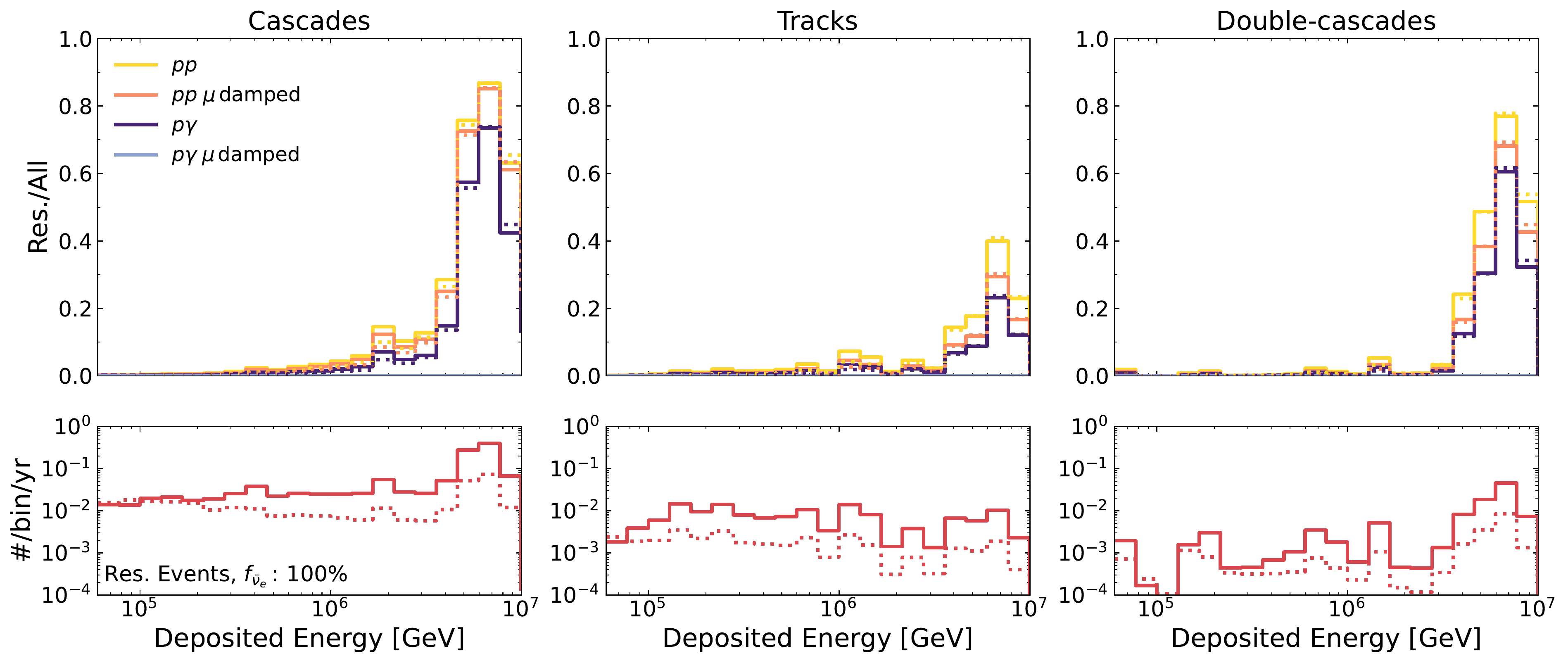}
    \caption{Top: the fractional contribution of GR to events of each morphology. Bottom: the event rate of GR events with IceCube exposure for each morphology. Here the $\bar{\nu}_e$ flux is normalized to the total high-energy neutrino flux, i.e. the event rate in a specific production scenario is to be multiplied by $f_{\bar{\nu}_e}$.  The solid lines and dotted lines correspond to the spectral index of $\gamma=2.37$ and $\gamma=2.87$, respectively.}
    \label{fig:src_event_dist}
\end{figure*}

\section{Analysis of GR events on statistical basis}
\label{sec:stats}

For the \textit{\textbf{statistical}} analysis, we perform a self-consistent analysis of the 3+1 flavor composition, by independently varying the four parameters $f_{\ell} \equiv f_{\nu_\ell }+ f_{\bar \nu_\ell}$ and $f_{\bar \nu_e}$, with $\ell = e,\mu, \tau$. We will compute likelihoods based on the expected number of events for the full range of astrophysical neutrino energies and morphologies, and include realistic misidentification rates.

\subsection{Analysis}

\begin{figure*}[tb!]
	\centering
	\hspace{-1cm}
	\includegraphics[width=0.5\textwidth]{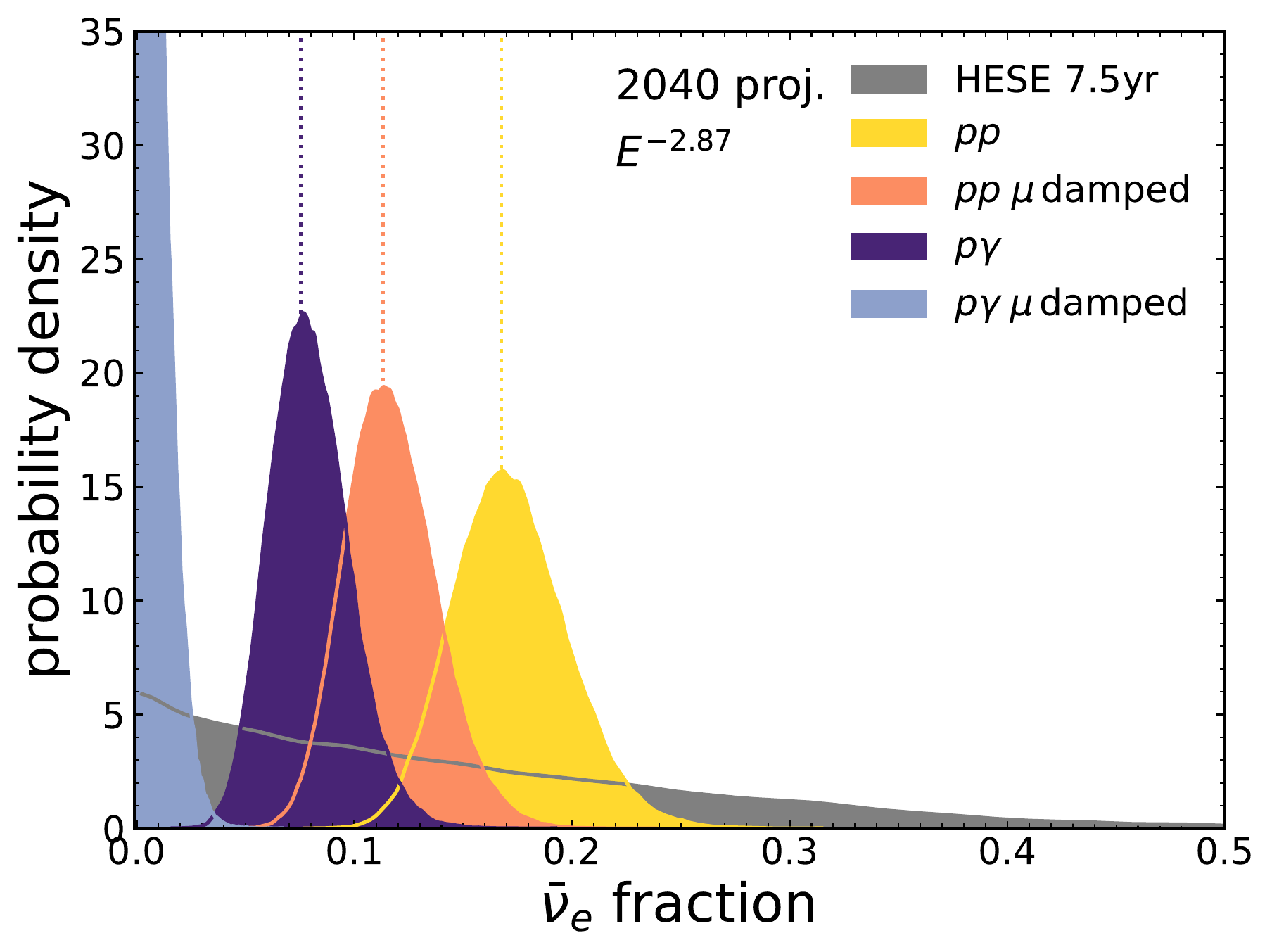}
	\includegraphics[width=0.5\textwidth]{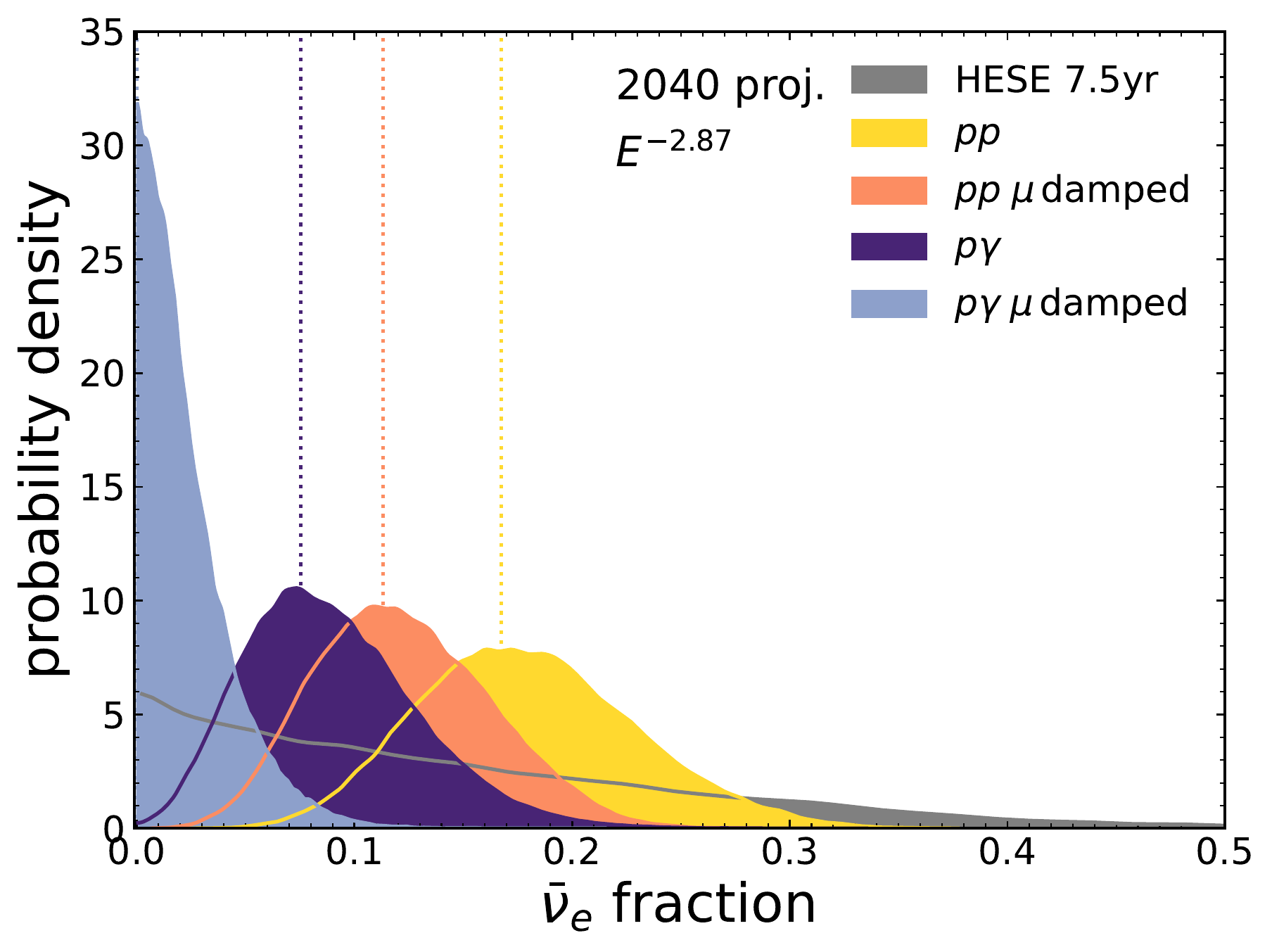}

    \caption{Posterior distribution of the fraction of $\bar{\nu}_e$ in total neutrino flux at Earth, $f_{\bar{\nu}_e}$, in the 4-flavor analysis. The gray region uses the HESE 7.5yr sample with no partially contained GR event. The colored ones employ the Asimov data generated for the combined exposure by 2040, assuming a specific production mechanism at the source. Non-detection of a GR event in the HESE sample results in a preference of 0\% $\bar{\nu}_e$. left: $E^{-2.37}$ neutrino spectrum. right: $E^{-2.87}$ neutrino spectrum.}
    \label{fig:prob_density_dist}
\end{figure*}

We will use the same framework for analyzing both current HESE (7.5 year) data \cite{IceCube:2020wum}, and the mock data that we generate for future telescopes.

Under the assumption of a given set of model parameters $\mathbf{\Theta}$, we consider cascades events as $N_{cas} = N^{GR}_{cas} +N^{NC,\nu/\bar{\nu}}_{cas} + N^{CC,\nu_e/\bar{\nu}_e}_{cas}$, track events $N_{tr} = N^{GR}_{tr} + N^{CC,\nu_\mu/\bar{\nu}_\mu}_{tr}$ and double-cascade events $N_{dc} = N^{GR}_{dc} + N^{CC,\nu_\tau/\bar{\nu}_\tau}_{db}$. 
We then perform a binned Bayesian analysis with the following likelihood function:
\begin{align}   
&  \mathcal{L} (\mathbf{\Theta})  = \prod_i \mathcal{L}_{cas,\,i}\prod_j \mathcal{L}_{tr,\,j}\prod_k \mathcal{L}_{db,\,k} \nonumber \\
 &= e^{-N_{cas} -N_{tr} - N_{dc} } \prod_i \frac{N^{N^{obs}_{cas,\,i}}_{cas,\,i}}{N^{obs}_{cas,\,i}!}
 \prod_j \frac{N^{N^{obs}_{tr,\,j}}_{tr,\,j}}{N^{obs}_{tr,\,j}!}
 \prod_k \frac{N^{N^{obs}_{dc,\,k}}_{dc,\,k}}{N^{obs}_{dc,\,k}!},
 \label{eq:likelihood_all}
\end{align}
$\mathbf{\Theta}$ represents the free parameters which will be discussed in this section and the superscript $obs$ is used to differentiate the observed number of events from the expected number of events.

For our statistical analysis, we use the binning scheme of the HESE diffuse flux analysis: the cascade and track events are binned in deposited energies and zenith angles while the double-cascades are binned in deposited energies and reconstructed lengths between the two cascades. Distinguishing cascade events and double-cascades events can be difficult depending on the length between two possible cascades. There are GR-induced events tagged as double-cascades in the HESE MC. These are likely to be misidentified cascades and have small reconstructed lengths. Since the misidentification is built into the the IceCube MC, this is implicitly taken into account in our likelihood evaluations. .

Since we are looking at a wider energy range and performing a full flavor analysis, more technical details would be needed to extend the MC to cover PEPE selection. In this section, we therefore restrict ourselves to the HESE selection. As previously mentioned, typical flavor analyses do not distinguish between neutrinos and anti-neutrinos.
We now turn to Bayesian analysis of the 3-flavor composition along with the fraction of $\bar{\nu}_e$. As $f_e + f_\mu + f_\tau = 1$, only two parameters are necessary to describe the flavor composition. To avoid bias in sampling, our priors are determined with the Haar measure~\cite{Haba:2000be} of the flavor ratio volume element $df_e \wedge df_\mu \wedge df_\tau$. We introduce flavor angles $\theta$ and $\psi$ where $(f_e,\ f_\mu,\ f_\tau) = (\mathrm{sin^2}\theta\, \mathrm{cos^2}\psi,\ \mathrm{sin^2}\theta \, \mathrm{sin^2}\psi,\ \mathrm{cos^2}\psi)$. Thus, the parameterization becomes $df_e \wedge df_\mu \wedge df_\tau = d\left(\mathrm{sin}^4 \theta \right) \wedge d\left(\mathrm{cos} 2\psi\right)$, and we impose uniform prior distributions $\mathrm{sin}^4 \theta \in \left[0,1\right]$ and $\mathrm{cos} 2\psi \in \left[-1,1\right]$.

In addition, we introduce a parameter $\kappa_{\bar{\nu}_e}$ with a uniformly distributed prior $\in \left[0,\,1\right]$, which represents the fraction of $\bar{\nu}_e$ in the total electron neutrino flux. The fraction of $\bar{\nu}_e$ in the total neutrino flux at Earth is $\kappa_{\bar{\nu}_e}f_{e}$. With two additional parameters describing the neutrino energy spectrum, i.e. the flux normalization $\phi_0$ and the spectral index $\gamma$, 5 free parameters are allowed to vary:  $\{\phi_0,\,\gamma,\,\kappa_{\bar{\nu}_e}, \theta,\psi \}$. 

In addition to the parameters characterizing the spectrum, there are other parameters accounting for the atmospheric background and systematic uncertainties in the IceCube HESE analysis \cite{IceCube:2020wum}. We fix these nuisance parameters to the best-fit values reported in their Table VI.1. We should note that Ref~\cite{IceCube:2020wum} did not vary the flavor composition, and instead assumed an equal contribution from each neutrino flavor. This could potentially impact the best fit locations of the nuisance parameters.

\subsection{Generation of mock future data}

 When computing the projected signal from our combination of future telescopes, we compute the events expected from a power-law spectrum and a 6-flavor neutrino composition, i.e. a composition with neutrino and antineutrino flavor ratios. This flavor composition is determined by 5 parameters: two sets of two flavor angles characterizing flavor compositions of neutrinos and antineutrinos, and one parameter $f_\nu$ representing the neutrino fraction of the total flux in neutrinos (as opposed to antineutrinos). i.e. we specify $\{\phi_0,\,\gamma,\,f_\nu, \theta_\nu,\psi_\nu ,\theta_{\bar{\nu}},\psi_{\bar{\nu}} \}$.

The top panels of Fig.~\ref{fig:src_event_dist} makes use of the HESE MC to show the expected ratios of GR events to the sum of GR events and DIS events for each morphology for the four pion decay neutrino production mechanisms. The bottom panels show the event rate expected with IceCube exposure in each energy bin if we were to assume 100\% of the neutrino flux is $\bar{\nu}_e$. The number of events for a given $\bar{\nu}_e$ fraction in a specific exposure can be obtained by multiplying the event rates by the effective livetime scaled by the real livetime and detector volume discussed in Sec.~\ref{sec:framework}. As expected, GR events dominate the event rates with energies around 6~PeV for cascades, while they remain subdominant for the track morphology even around the resonant energy. The event rates of double-cascade GR events are comparable to that of tracks. Those are however likely to be misidentified, as stated earlier.  The spectral index also notably affects the event rates. There is a factor of 5.4 difference between the hard and soft spectrum assumptions in the event rate at the energy bin around 6~PeV.

\subsection{Posterior distributions}

\begin{figure}[htb!]
    \centering
    \includegraphics[width=0.5\textwidth]{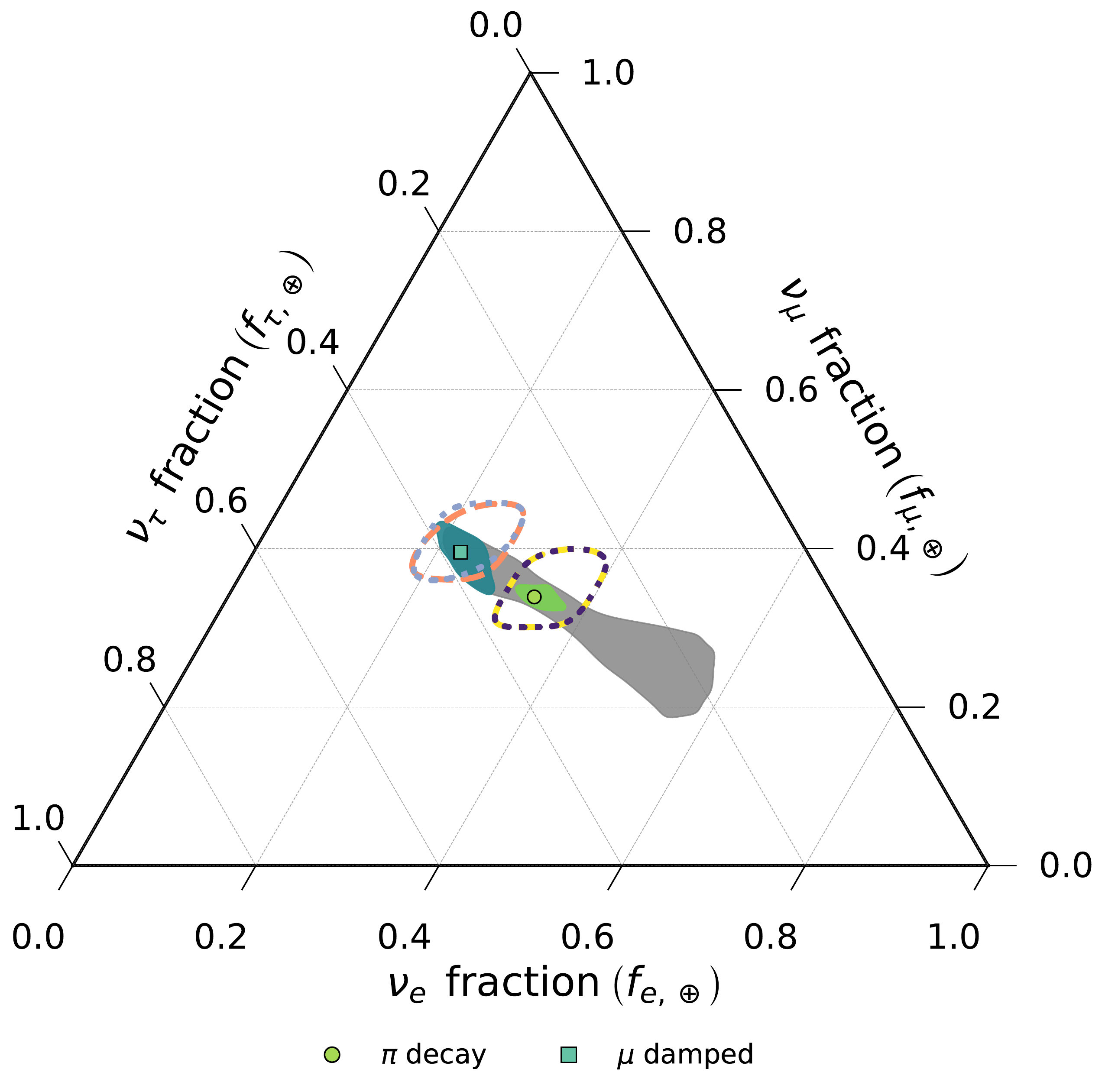}
    \caption{90\% credible contours of the 3-flavor compositions for pion decay and muon damped scenarios projected to 2040. Dashed contours correspond to $pp$ and dotted contours correspond to $p\gamma$. The colors match the ones in Fig.~\ref{fig:prob_density_dist}. The two processes are degenerated as expected while the pion decay and muon damped scenarios will be well differentiated in the future 3-flavor analysis. With the use of NuFIT~5.1 oscillation results, the shaded green regions show the oscillation allowed regions for the two scenarios and the gray region is the standard oscillation allowed region.}
    \label{fig:triangle_2040}
\end{figure}

The posterior distributions of $f_{\bar{\nu}_e}$, which is transformed from $\kappa_{\bar{\nu}_e}$ and $f_e$ using the flavor angles, are shown in Fig.~\ref{fig:prob_density_dist}. They are computed using the HESE data and the Asimov data for the combined exposure by 2040 considering the four neutrino production mechanisms. 
The HESE result constrains the $\bar{\nu}_e$ fraction to $\lesssim 0.3$ with 90\% credibility, with most probable value at 0.002. This is consistent with the non-observation of a GR event in HESE and the expectation that high-energy neutrino sources generate $f_{\bar{\nu}_e} \lesssim 0.2$ at Earth, consistent with all scenarios discussed here. By 2040, the widths of the posterior distributions shrink markedly compared with HESE 7.5yr data, making it possible to differentiate production scenarios. In particular, the scenarios that are degenerate in a 3-flavor analysis, i.e. $pp$ vs $p\gamma$ and their muon damped scenarios, can be resolved with future observations. With the hard spectrum, $pp$ vs $p\gamma$ can be distinguished at more than 3$\sigma$ significance. 
 Even considering the conservative assumption for the spectrum,  $pp$ and $p\gamma$ can still be separated with a significance above 2$\sigma$. The two production mechanisms in the muon-damped scenarios can be differentiated at more than 3$\sigma$ significance.

Our 4-flavor analysis still has the power to distinguish the pion decay from the muon-damped scenarios. The posterior contours of the 3-flavor composition in the analysis are displayed in Fig.~\ref{fig:triangle_2040}. Two 90\% contours are well separated from each other by 2040, comparable to the results in~\cite{Song:2020nfh} which performed a 3-flavor analysis without distinguishing neutrinos and antineutrinos. 
 The joint posterior distributions for the parameters used in the 4-flavor analysis is shown in Fig.~\ref{fig:2040_pg_gamma2.37_corner_plot} in Appendix~\ref{sec:appendix}. The results can be further improved in the future with better event reconstruction such as event identification and energy reconstruction, and inclusion of the PEPE selection, as discussed in Sec.~\ref{sec:GR_ID}.

\begin{table*}[htb!]
 \centering
    \begin{tabularx}{\textwidth}{YYYYYY}
    \hline
    \multirow{2}{*}{Analysis} & \multirow{2}{*}{Spectrum}  & $pp$ from $p\gamma$  & $p\gamma$ from $pp$  &  $pp$ from $p\gamma$   &  $p\gamma$ from $pp$ \\
    &  & $\pi$ decay & $\pi$ decay & $\mu$ damped & $\mu$ damped \\
    \hline

    HESE event-wise & soft & 1.6$\sigma$ & 1.4$\sigma$ & $> 5\sigma$ & 0.7$\sigma$ \\
         & hard & 3.8$\sigma$ & 3.3$\sigma$ & $> 5\sigma$&  6.0$\sigma$   \\
    PEPE event-wise & soft & 2.3$\sigma$ & 2.0$\sigma$ & $> 5\sigma$ &  1.4$\sigma$  \\
         & hard & 5.3$\sigma$  & 4.7$\sigma$ & $> 5\sigma$ &  6.9$\sigma$ \\  
         
    \hline
    HESE Bayesian & soft & 2.6$\sigma$ & 2.1$\sigma$ &  3.5$\sigma$ & 3.1$\sigma$ \\
     & hard & 4.4$\sigma$ & 3.9$\sigma$ & 6.3$\sigma$ & 6.5$\sigma$ \\
    \hline    
    \end{tabularx}
    \caption{The significance levels we expect to distinguish one scenario from the other by 2040. We consider two neutrino production mechanisms, pion decay and muon-damped pion decay, which each produce the same 3-flavor ratio for $pp$ and $p\gamma$ sources, but are distinguishable via the GR. In the first four rows, we assume event-wise identification of GR, and infer the probability to distinguish the true production scenario from the alternate using the HESE event selection or PEPE event selection with a frequentist method (Sec.~\ref{sec:GR_ID}). In the last two rows, we relax this assumption and assume GR can only be distinguished statistically. We adopt HESE event selection and infer the significance with the Bayesian approach (Sec.~\ref{sec:stats}). Soft and hard neutrino spectra are analyzed in each study with the spectral index $\gamma=2.87$ and $\gamma=2.37$,  respectively. 
    }
    \label{tab:significance}   
\end{table*}

\section{Impact of mixture and more complex conditions} 
\label{sec:combination}

In previous sections we focused on the high-energy astrophysical neutrinos assuming a single production mechanism. Realistically, it is reasonable to expect that there exists more than one source population, and different neutrino production mechanisms contribute to the total neutrino flux. This scenario has been examined in~\cite{Bustamante:2019sdb} in a 3-flavor study allowing arbitrary neutrino ratio at the source, and in~\cite{Song:2020nfh} for a combination of the benchmark scenarios. The signature of GR events from mixed $pp$ and $p\gamma$ sources is also studied in~\cite{Bhattacharya:2011qu}. 

We will focus on scenarios that lead to the same 3-flavor compositions, but different neutrino-to-antineutrino ratios.\footnote{Differentiating e.g. the pion decay scenario $(1:2:0)_s$ from the muon-damped case $(0:1:0)_s$ can be readily done using flavor alone, see, for example \cite{Song:2020nfh}.} We start with the $pp$ and $p\gamma$ scenarios, before moving onto a more general case, varying the $\pi^+$-to-$\pi^-$ ratio. Such scenarios could occur e.g. due to secondary production after $pp$ or $p\gamma$ interactions.

\subsection {Mixture of $pp$ and $p\gamma$}
First, we examine a mixture of general $pp$ and $p\gamma$ production mechanisms where we expect a $\bar{\nu}_e$ fraction resting in the ranges given by the two scenarios in Tab.~\ref{tab:flavor_ratio}. We introduce the parameter $f_{pp}$ ($= 1-f_{p\gamma}$) to indicate the fraction of $pp$ interactions in a mixed contribution. Following the neutrino production introduced in Sec.~\ref{sec:source}, after neutrino oscillation, we obtain the relations between $f_{\bar{\nu}_e}$ and $f_{pp}$ at Earth: $f_{\bar{\nu}_e} = 0.092f_{pp} + 0.075 $ for regular pion decay, and $f_{\bar{\nu}_e} = 0.113f_{pp}$ in the muon-damped case. We generate Asimov data with the neutrino and antineutrino compositions given by a specific $f_{pp}$. When computing the posteriors, we again let the spectral index and flux normalizations vary as free parameters while allowing a uniform distribution of $f_{pp}\in \left[0,1\right]$, but we assume that the 3-flavor composition is independently determined, resulting in 3 free parameters, i.e. $\{\phi_0,\,\gamma,\,f_{pp}\}$. 

\begin{figure*}
    \centering
    \hspace{-1cm}
    \includegraphics[width=0.5\textwidth]{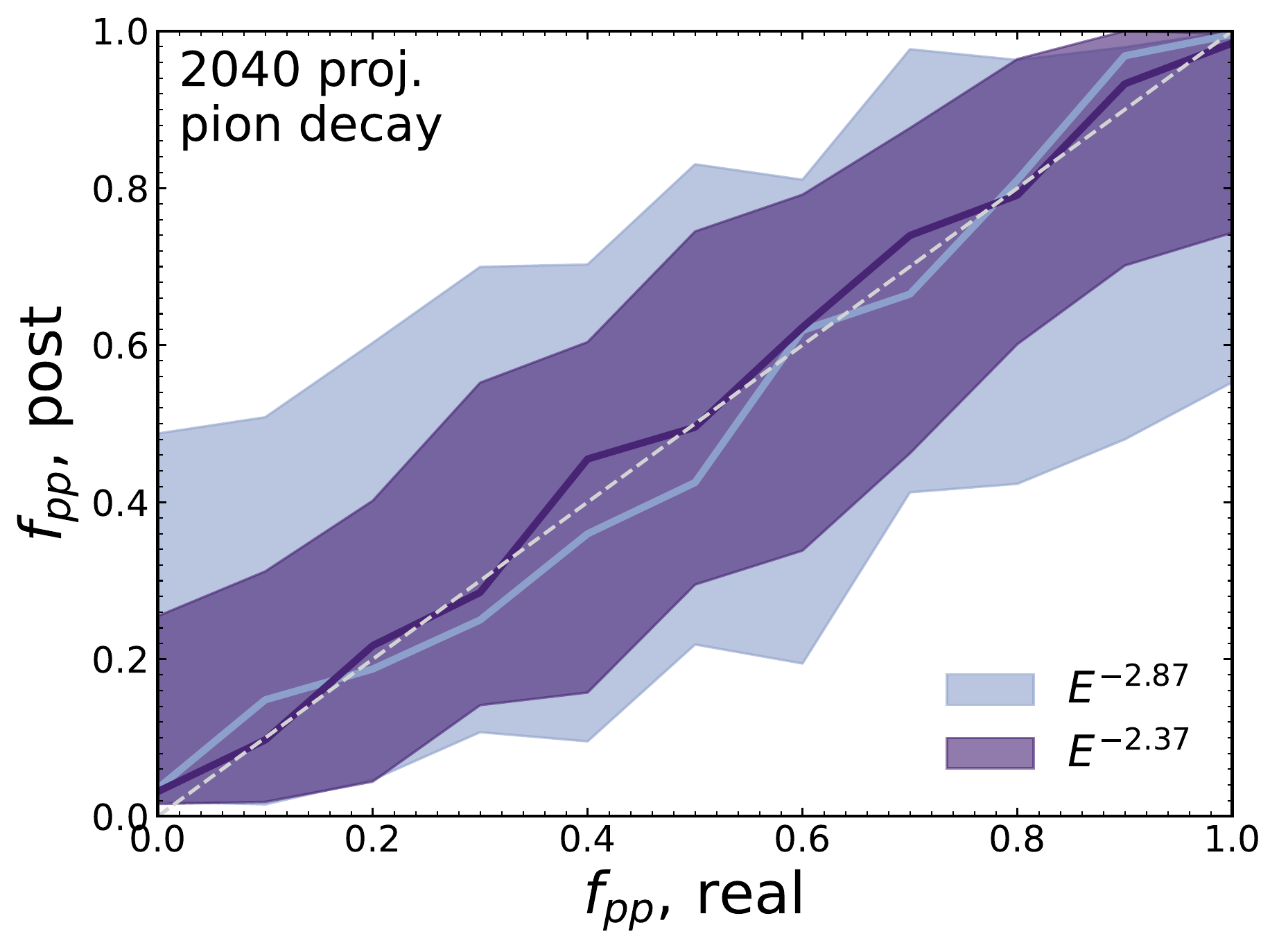}
	\includegraphics[width=0.5\textwidth]{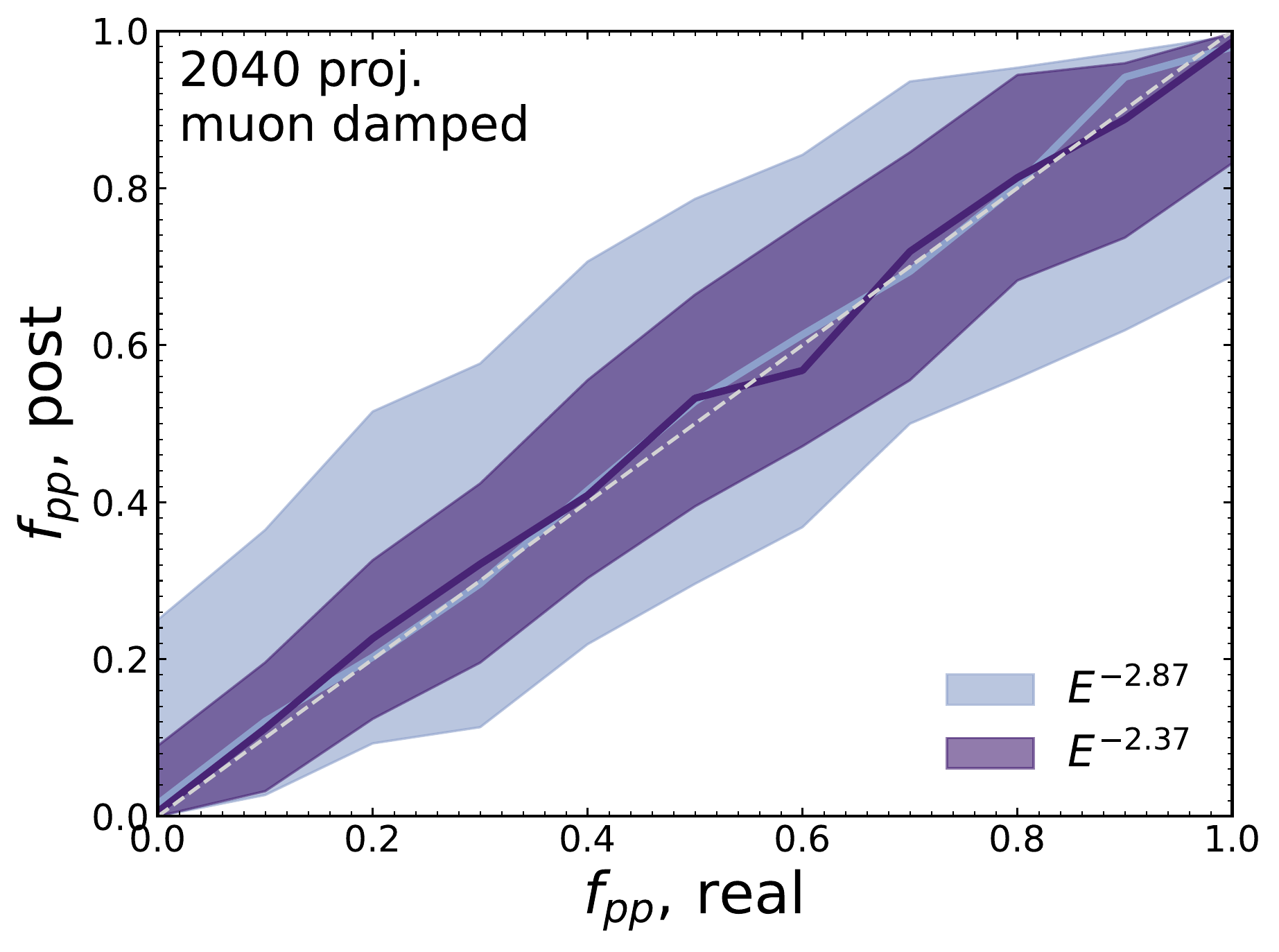}
    \caption{68\% posterior credible intervals as a function of the real $f_{pp}$ values in the Asimov data for pion decay (left) and muon damped pion decay (right) scenarios by 2040 with the soft and hard spectrum assumptions. The central lines show the most probable values.}
    \label{fig:fpp_ratio}
\end{figure*}

\begin{figure*}
\hspace{-1cm}
    	\includegraphics[width=0.5\textwidth]{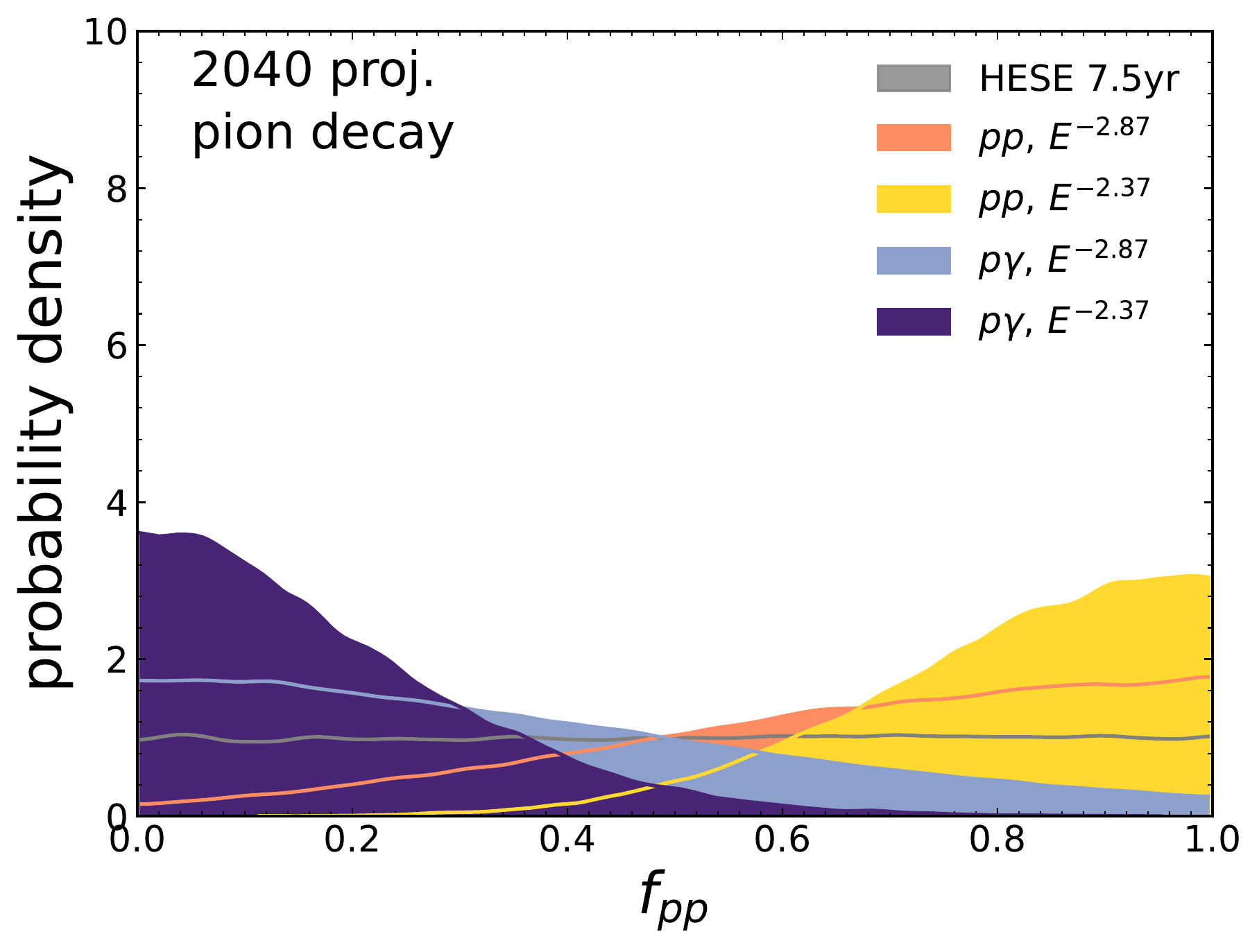}
    \includegraphics[width=0.5\textwidth]{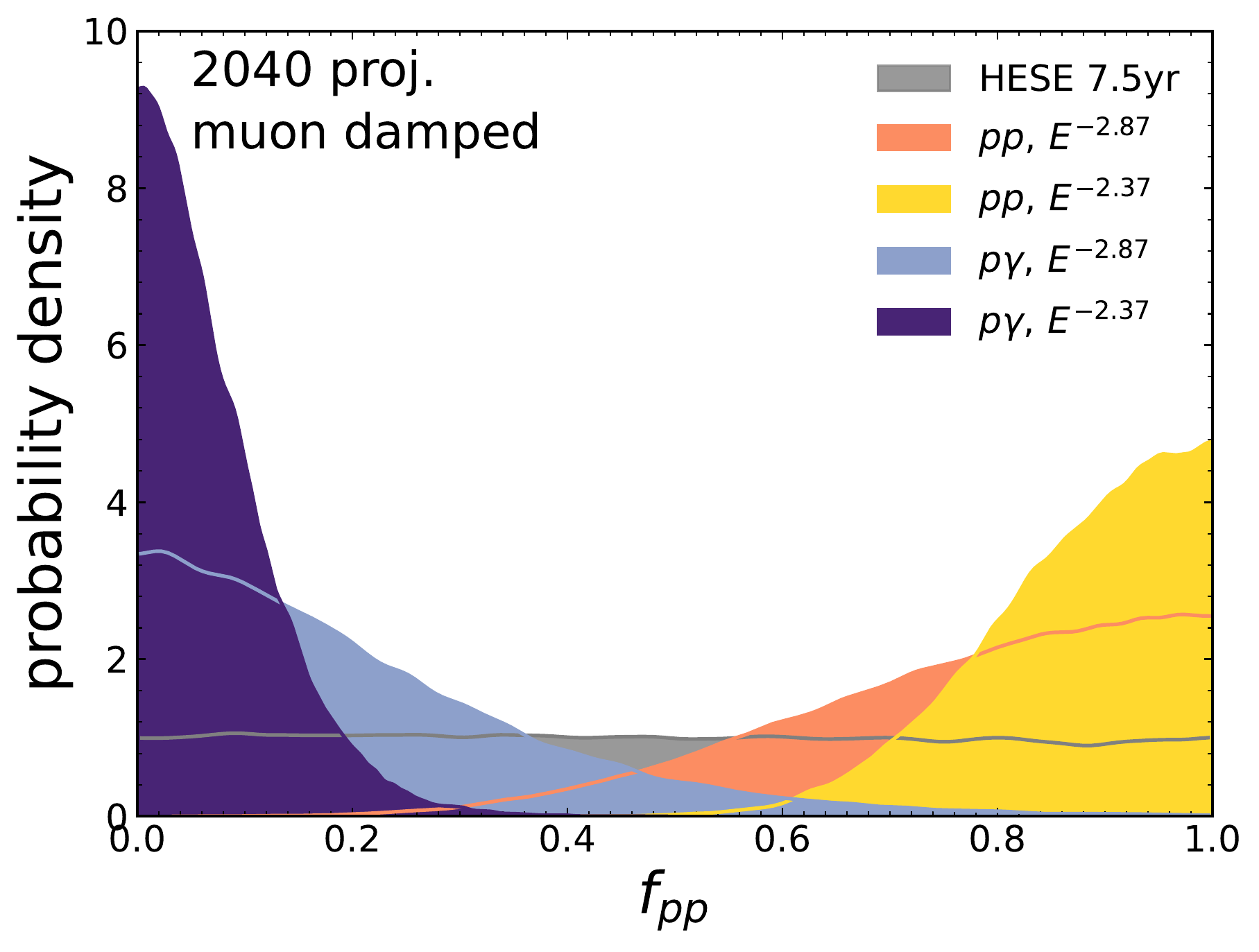}
    \caption{Posterior distributions of $f_{pp}$ for pion decay (left) and muon damped pion decay (right) scenarios for 2040, as compared with the results using the HESE 7.5yr data. Each panel shows the $f_{pp}$ posterior distributions when the true flux is from a pure $pp$ process or a pure $p\gamma$ process under the soft and hard spectrum assumptions.  }
    \label{fig:fpp}
\end{figure*}

Fig.~\ref{fig:fpp_ratio} shows the 68\% posterior credible intervals of $f_{pp}$ for pion decay and muon-damped scenarios by 2040. Due to a steeper slope in the $f_{\bar{\nu}_e}-f_{pp}$ relation, future neutrino telescopes have better sensitivity to pin down the contribution from $pp$ and $p\gamma$ separately in the muon damped scenario than in the pion decay scenario. With the soft spectrum assumption, the uncertainty band of the pion decay scenario is $\sim 0.4 - 0.61$ and is improved to $\sim 0.24-0.45$ with the harder spectrum assumption. For the muon damped scenarios, it becomes $\sim 0.25-0.49$ with the soft spectrum and $\sim 0.08-0.29$ with the hard spectrum. The posterior distributions of $f_{pp}$ can be found in Fig.~\ref{fig:fpp}, for the specific cases where true values of $f_{pp}=0$ and $f_{pp}=1$. The analysis using the HESE 7.5yr data yields quite flat distributions, indicating the current high-energy neutrino measurement has no power to tell the mixture of the neutrino flux from $pp$ and $p\gamma$. The discrimination power is vastly improved in the future, with more data available in larger neutrino telescopes.

\subsection {Mixture of $\pi^+$ and $\pi^-$}

The neutrino flavor composition in each scenario may not exactly follow the ideal flavor ratios we have examined so far. Depending on the energies of the injecting coscmic rays and targets, subdominant interactions can perturb the common approximation of equally produced $\pi^+$, $\pi^-$ and $\pi^0$ for $pp$ interactions, and the purely $\Delta^+$ resonant production in $p\gamma$ interactions.  

From detailed MC simulations for photohadronic processes, the ratio of $\pi^+/\pi^-$ deviates from the ideal values as a consequence of direct pion production, higher resonances, multipion production, etc~\cite{Mucke:1999yb}. For hadronuclear processes, MC simulations using high-energy hadronic interaction models widely used for cosmic ray interactions such as SYBILL and QGSJET~\cite{Riehn:2017mfm,Ostapchenko:2010vb} predict that valence quark scattering and hadronization take a large momentum fraction. This leads to more $\pi^+$ given more $u$ valence quarks in protons, rather than equal number of $\pi^+$ and $\pi^-$. This is further complicated by the fact that the composition of cosmic rays is unresolved. Apart from the subdominant interactions, a fraction of heavy chemical elements at the highest energies is suggested in cosmic ray measurements~\cite{PierreAuger:2016use,TelescopeArray:2018bep,PierreAuger:2022atd}, which can also interact and produce neutrinos with different $\pi^+$ to $\pi^-$ ratios. Another proposed scenario is that if the source environment is optically thick for photohadronic interactions, a predominant neutrino flux from $\pi^-$ decay can also be expected for sources where protons are magnetically confined leading to the dominant charged pion production via $n + \gamma \rightarrow \Delta^0 \rightarrow \pi^- + p$~\cite{Baerwald:2013pu}. Part of the scenarios which may cause the deviation from ideal $pp$ and $p\gamma$ interactions are discussed in e.g.~\cite{Biehl:2016psj}. 

Since an exact model of the neutrino spectrum relies on unknown information about the source composition and properties, we cannot produce a realistic model. Hence, we will model the range of pion production scenarios in the ratio of $\pi^+$ / $\pi^-$. GR events provide us with a powerful tool to diagnose the asymmetry of $\pi^+$ and $\pi^-$ in such processes, which then reveals more information about interactions in neutrino sources. We introduce the parameter $f_{\pi^+}$, which represents the fraction of $\pi^+$ in the total number of charged pions produced in the source. The relations between $f_{\bar{\nu}_e}$ at Earth and $f_{\pi^+}$ at the source are $f_{\bar{\nu}_e} = -0.184f_{\pi^+} + 0.259 $ for direct pion decay, and $f_{\bar{\nu}_e} = -0.226f_{\pi^+} + 0.226$ in the muon damped case. The fraction of $\bar{\nu}_e$ ranges from 0.08 to 0.26 as $f_{\pi^+}$ varies from 1 to 0. This becomes 0 to 0.23 when muon decay is suppressed. We again impose uniform prior on $f_{\pi^+}$ $\in \left[0,1\right]$ and keep the spectral index and flux normalization as free parameters, i.e. we vary $\{\phi_0,\,\gamma,\,f_{\pi^+}\}$.

\begin{figure*}[t!]
    \centering
	\hspace{-1cm}
	\includegraphics[width=0.5\textwidth]{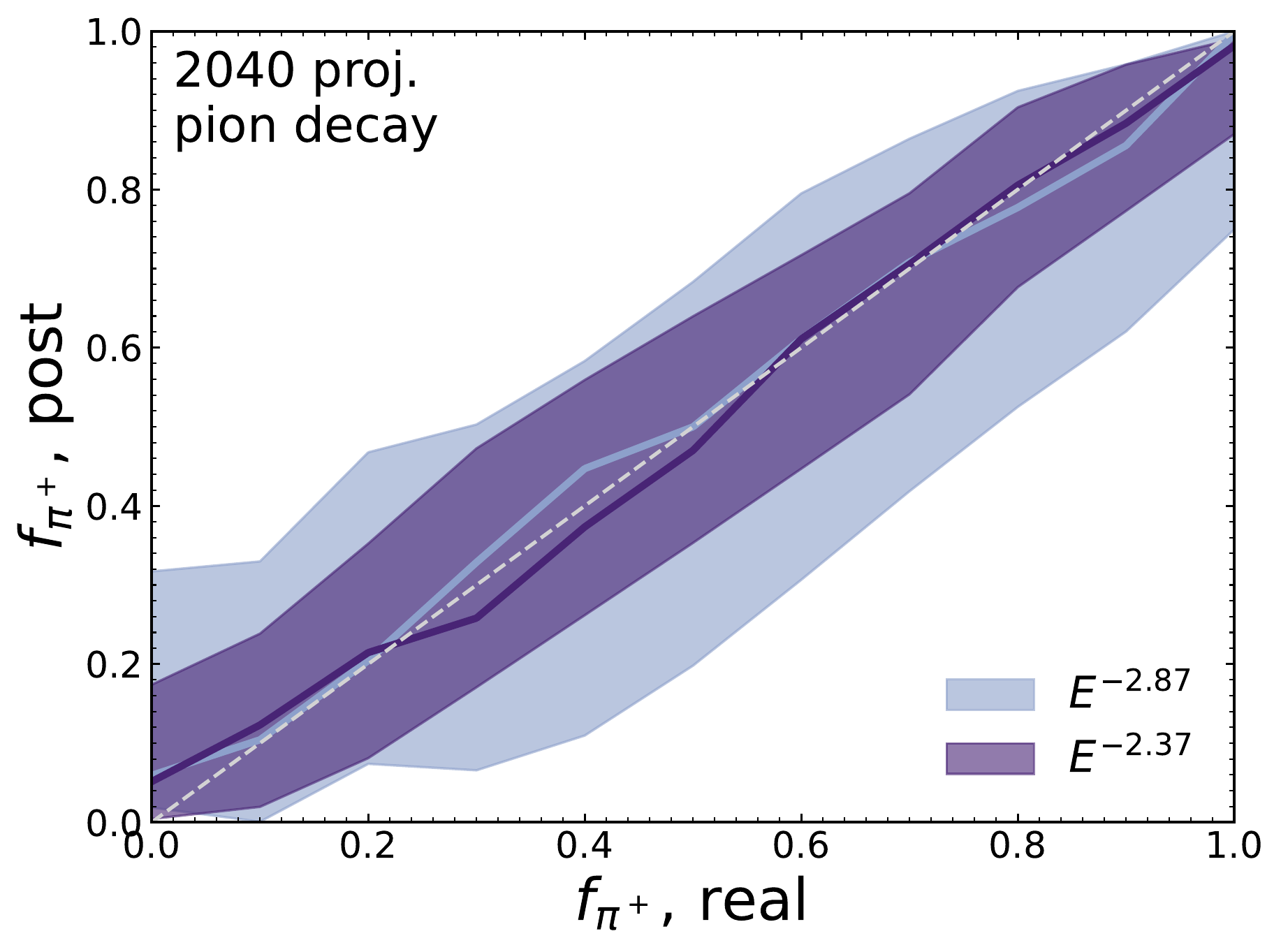}
    \includegraphics[width=0.5\textwidth]{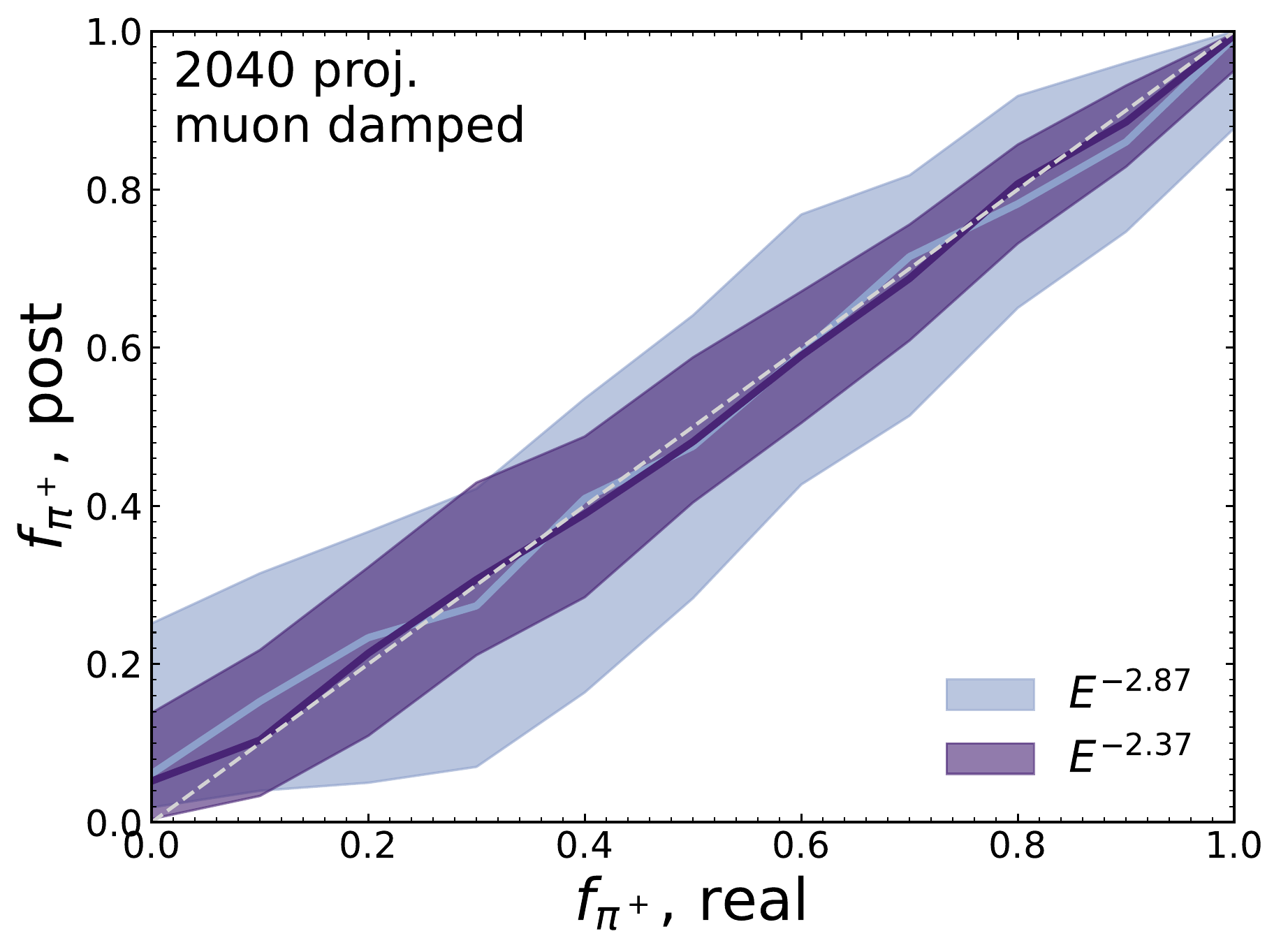}
    \caption{68\% posterior credible intervals as a function of the real $f_{\pi^+}$ values in the Asimov data for pion decay (left) and muon damped pion decay (right) scenarios by 2040 with the soft and hard spectrum assumptions. The central lines show the most probable values.}
    \label{fig:pi_ratio}
\end{figure*}

\begin{figure*}
\hspace{-1cm}
    	\includegraphics[width=0.5\textwidth]{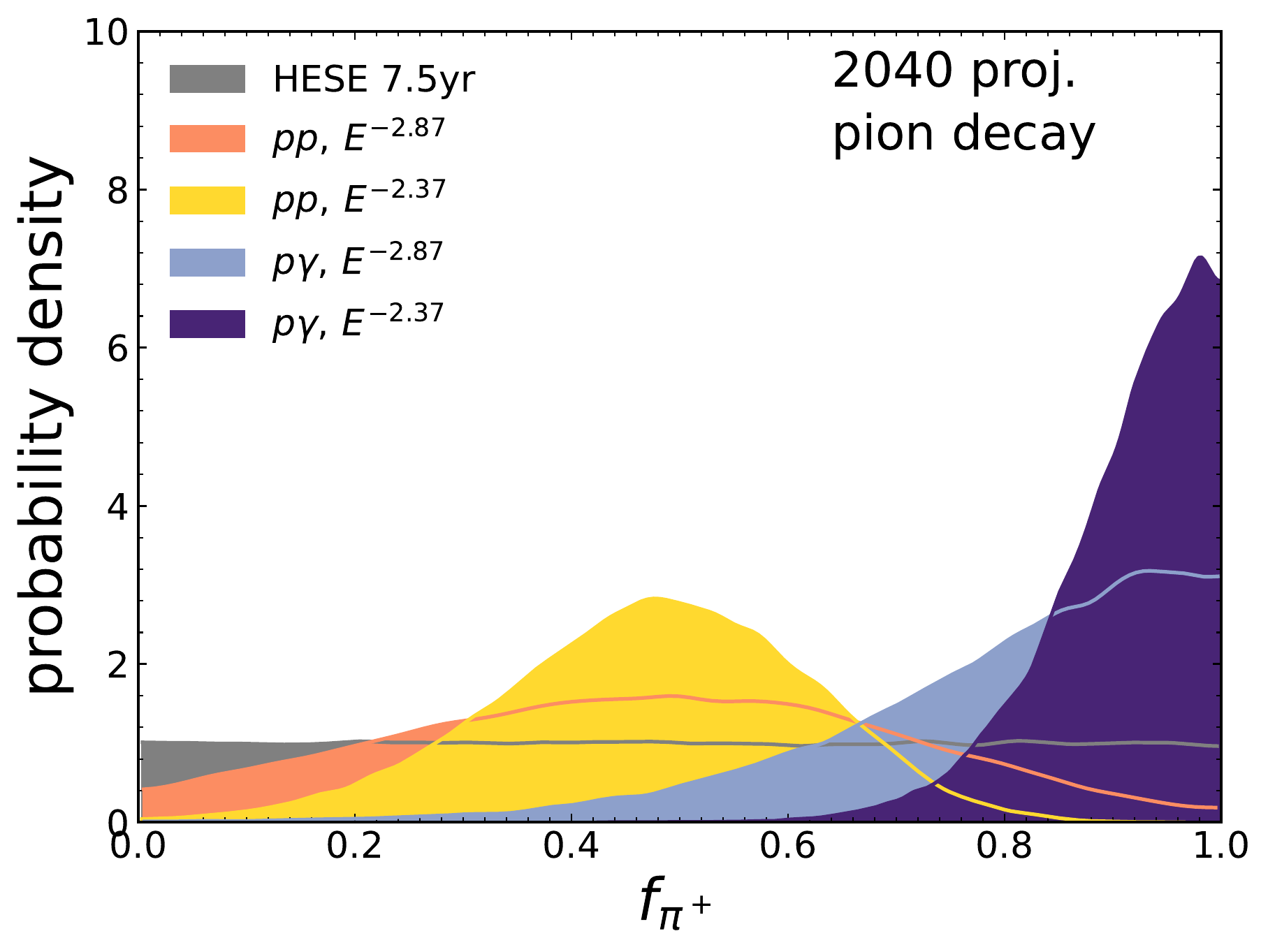}
    \includegraphics[width=0.5\textwidth]{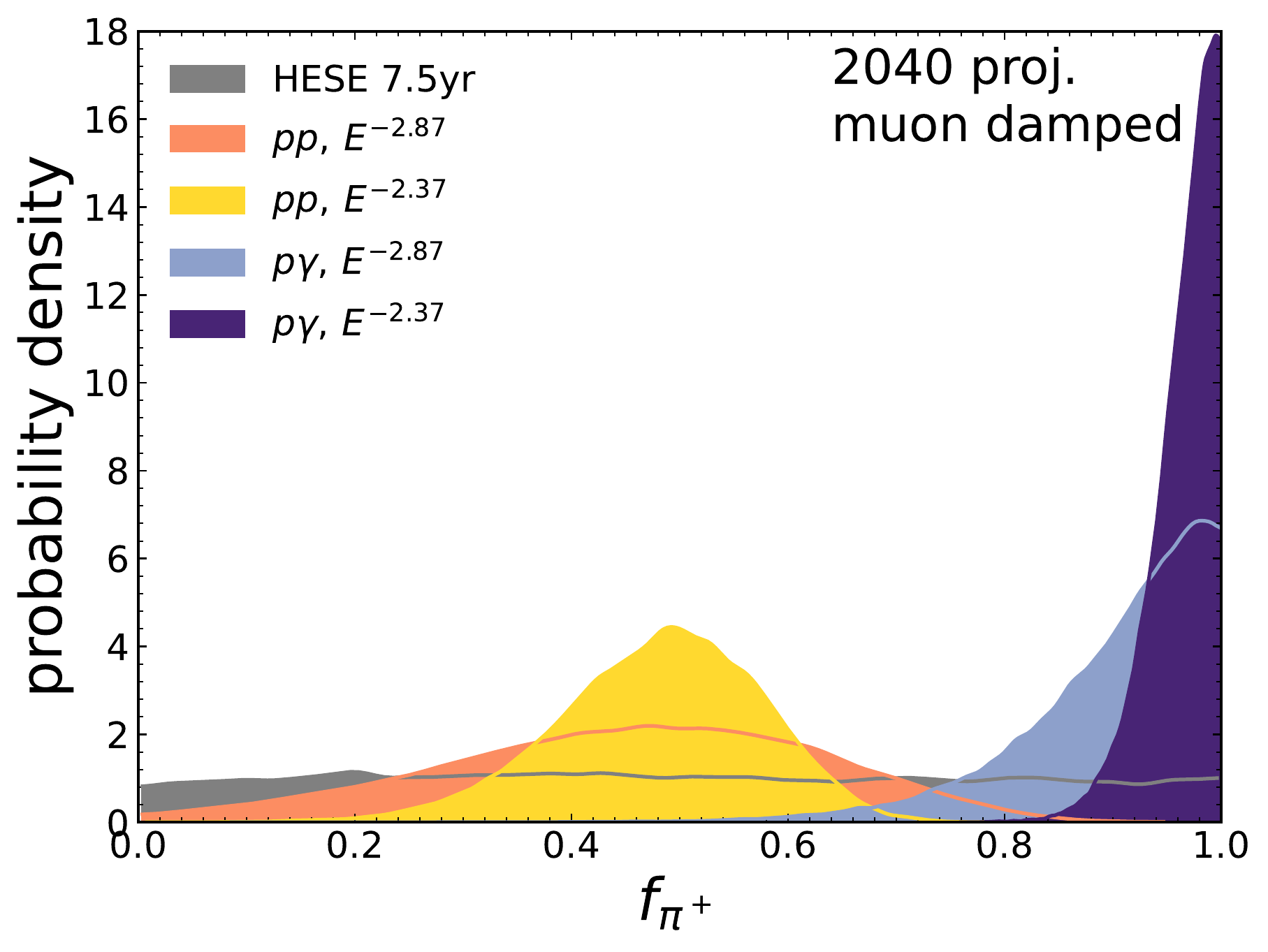}
    \caption{Posterior distributions of $f_{\pi^+}$ for pion decay (left) and muon damped pion decay (right) scenarios for 2040, as compared with the results using the HESE 7.5yr data. Each panel shows the $f_{\pi^+}$ posterior distributions when the true flux is from a pure $pp$ process or a pure $p\gamma$ process under the soft and hard spectrum assumptions. }
    \label{fig:fpi}
\end{figure*}

Fig.~\ref{fig:pi_ratio} shows the 68\% posterior bands of $f_{\pi^+}$ by 2040 for pion decay and muon-damped sources. Similar to the discussion on the mixed $pp$ and $p\gamma$ flux, $f_{\pi^+}$ is better determined in the muon-damped scenario owing to the slope of the $f_{\bar{\nu}_e}-f_{\pi^+}$ relation. For the pion decay scenario, the posterior uncertainty band has a width ranging from $\sim 0.25-0.49$ with the soft spectrum and $\sim 0.12-0.3$ with the hard spectrum. For the muon damped scenario, these values decreases to $\sim 0.12-0.37$ with the soft spectrum and $\sim 0.05-0.22$ with the hard spectrum. The posterior distributions of $f_{\pi^+}$ can be found in Fig.~\ref{fig:fpi}, in the specific cases where the true values of $f_{\pi^+}=0$ and $f_{\pi^+}=1$. At present, the HESE 7.5yr data  does not tell us much about the asymmetry of pions in the source: the posterior is flat over the entire range due to the non-observation of GR events in the sample.  Remarkable improvement is expected in the advent of future neutrino telescopes, as showcased using the benchmark (idealized) scenarios ($pp$: $f_{\pi^+} = 0.5$ versus $p\gamma$: $f_{\pi^+} =1$) in Fig. \ref{fig:fpi}.

These results use the current HESE selection and reconstruction methods. Improvements to such methods in the future will of course help particle-wise identification and fine-graining of the statistical approach.

\section{Conclusion}
\label{sec:conclusion}
High-energy astrophysical neutrinos have long been regarded as a unique probe of their distant origin, thanks to their capability of traveling unimpededly to the Earth. Source information is encoded in both their flavor and energy spectrum. While IceCube marked a great success in pinning down the neutrino flavor composition at Earth, upcoming neutrino telescopes will be able to precisely determine the neutrino production mechanisms, mostly notably distinguishing pion decay and muon-damped scenarios. However, ambiguity remains in the asymmetry of pions at the source. Although $pp$ and $p\gamma$ interactions predict a different amount of $\pi^+$ and $\pi^-$, the flavor composition, adding up neutrinos and antineutrinos, remains the same.

The observation of a GR event in IceCube shines light on the possibility of breaking this degeneracy. As the cross section of $\bar{\nu}_e$ near the resonant energy is vastly enhanced, the detection of GR events facilitates the determination of the fraction of $\bar{\nu}_e$, and antineutrinos in general thanks to oscillation. Starting from the unique signatures of GR events, we have studied the fraction of $\bar{\nu}_e$ assuming GR can be distinguished on event-by-event basis. The uncertainties on $f_{\bar{\nu}_e}$ are constrained to be between $5\%-8\%$ by 2040 combining the exposure of future neutrino telescopes, depending on the assumption of the high-energy neutrino spectrum. Conservatively, if GR events are indistinguishable from neutrino-nucleon DIS events, we also performed a statistical analysis on the 3-flavor composition plus the fraction of $\bar{\nu}_e$. In both analyses, we found that the previously degenerate $pp$ and $p\gamma$ interactions can be well separated by 2040, presuming that we can combine data from the full contingent of upcoming Cherenkov neutrino telescopes.

Realistic sources may involve more than one production mechanism, e.g. both $pp$ and $p\gamma$ and even with a specific production mechanism, Monte Carlo simulations suggest that the amount of $\pi^+$ and $\pi^-$ may deviate from the standard values in theoretical predictions. We have accounted for these possibilities by studying the fraction of high-energy neutrinos from $pp$ interactions, and the fraction of $\pi^+$ in the total charged pion population at the source. We find that by 2040, the contribution from $pp$ sources can be determined with a $0.25-0.61$ uncertainty with a soft spectrum, falling to $0.08-0.45$ with a hard spectrum. For a mixture of $\pi^+$ and $\pi^-$ at the source, the fraction of $\pi^+$ can be constrained with an uncertainty between $0.12$ and $0.49$ for a soft spectrum, and $0.05-0.37$, for a hard spectrum.

Beyond identifying high-energy neutrino sources and production mechanisms, the revolutionary advancement in  determining the $\bar{\nu}_e$ flux with GR also paves the way for disentangling the effects of potential new physics. Ref.~\cite{Bustamante:2020niz} set limits on neutrino lifetime by investigating the observed GR event assuming inverted neutrino mass ordering. More generally, new physics could introduce asymmetry between neutrinos and antineutrinos, while leaving 3-flavor composition unchanged. Such scenarios include the decay of asymmetric dark matter (ADM) which produces more antineutrinos than neutrinos~\cite{feldstein2010discovering,Fukuda:2014xqa,Zhao:2014nsa}, or ADM that preferably interacts with neutrinos or antineutrinos, which degrades the energy of neutrinos. If neutrinos are Majorana, a transition neutrino magnetic moment allows the interconversion between neutrinos and antineutrinos~\cite{Akhmedov:2003fu,Ando:2003is,Jana:2022tsa,Kopp:2022cug}. This is particularly important for a $p\gamma$ muon-damped source. With standard oscillation we expect no $\bar{\nu}_e$ at Earth. However, the source and intergalactic magnetic field may facilitate the conversion from $\nu_\mu$ to $\bar{\nu}_e$, producing GR events. We leave the study of new physics scenarios for future work.

Apart from the detection in Cherenkov neutrino telescopes, GR events can also be probed in TAMBO-like experiments~\cite{Romero-Wolf:2020pzh}, where Earth-skimming $\bar{\nu}_e$ interacts in the mountain, producing $\tau$ which exits and decays into air shower signals~\cite{Huang:2019hgs}. The expected GR event rate per year in TAMBO is 0.46$f_{\bar{\nu}_e}$ assuming the soft spectrum and increases to 2.5$f_{\bar{\nu}_e}$ with the hard spectrum. For a $\tau$ extensive air shower experiment,  GR events cannot be distinguished from $\nu_\tau$ CC events on an even-by-event basis as both signals are induced by $\tau$ decay.  The interaction length of GR in rock is $\sim 26$~km, and the decay length of $\tau$ is $\sim 50 (E_\tau/\mathrm{PeV})$~m. Consequently, mountains or rocks with a thickness of several kilometers make the best targets for GR detection. Future extensive air shower experiments will increase the exposure of GR detection.

It would also be beneficial to combine the GR analysis with a spectral analysis. The source information extracted from GR only applies around the resonant energy of a few PeV, while the dominant production mechanisms at different energy ranges may differ. For example the transition to the muon-damped scenario could happen from low to high energies. The exploration of these scenarios is interesting and important for future work.

Current data are consistent with the standard production mechanisms discussed here. However, with the advent of half a dozen new Cherenkov neutrino telescopes around the globe, it is clear that, the value of the $W$ mass is placed just well enough so that the origin of high-energy neutrinos cannot hide for long. 

\begin{acknowledgments}
The authors would like to thank Tianlu Yuan for useful discussions and Donglian Xu for correspondence on TRIDENT experiment.

QL and ACV are supported by the Arthur B. McDonald Canadian Astroparticle Physics Research Institute, with equipment funded by the Canada Foundation for Innovation and the Province of Ontario, and housed at the Queen’s Centre for Advanced Computing. Research at Perimeter Institute is supported by the Government of Canada through the Department of Innovation, Science, and Economic Development, and by the Province of Ontario. NS is supported by the National Natural Science Foundation of China (NSFC) Project No. 12047503. NS also acknowledges the UK Science and Technology Facilities Council for support through the Quantum Sensors for the Hidden Sector collaboration under the grant ST/T006145/1. ACV is also supported by NSERC, and the province of Ontario via an Early Researcher Award. 

\end{acknowledgments}

\bibliography{ref2}

\newpage

\onecolumngrid
\appendix

\section{Joint posterior distribution from 4-flavor analysis}
\label{sec:appendix}
\begin{figure*}[htb!]
    \centering
    \includegraphics[width=0.88\textwidth]{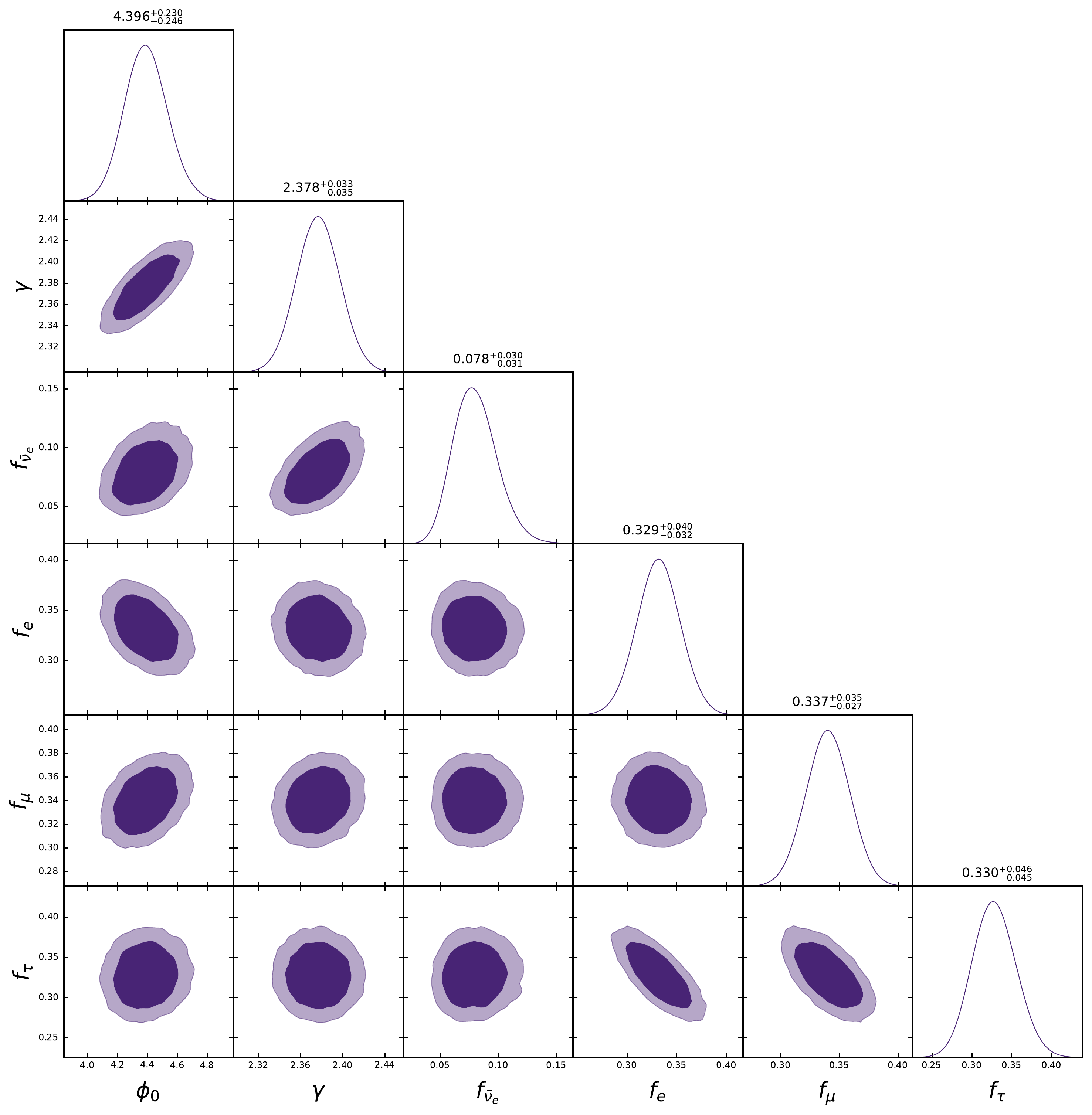}
    \caption{Joint posterior distributions for the astrophysical flux parameters in our 4-flavor analysis for the combined exposure to 2040. The 3 flavor fractions are transformed from the unbiased parameters, i.e. flavor angles used for sampling and $f_{\bar{\nu}_e}$ is obtained from  $f_{\bar{\nu}_e}=\kappa_{\bar{\nu}_e}f_e$ . The measured through-going muon spectrum is assumed with the assumption of the $p\gamma$ scenario.}
    \label{fig:2040_pg_gamma2.37_corner_plot}
\end{figure*}

\end{document}